\documentclass[draft]{agujournal2019}
\usepackage[utf8]{inputenc}
\usepackage{textgreek}
\usepackage{url} 
\usepackage{lineno}
\usepackage[inline]{trackchanges} 
\usepackage{soul}
\usepackage{amsmath}
\usepackage{float}
\usepackage{booktabs}
\usepackage{subcaption}
\usepackage{multirow}
\usepackage{subcaption}
\usepackage{mathtools}
\usepackage[section]{placeins}
\usepackage{rotating}
\usepackage{adjustbox}

\usepackage{lipsum} 

\usepackage{xcolor}
\usepackage[breakable]{tcolorbox}

\draftfalse
\usepackage{amssymb}
\usepackage{wrapfig}

\journalname{AGU}

\begin{document}

\title{Fast and Stable Nonlinear Emulation of TIEGCM Using an Autoregressive SINDy Framework}

\authors{Daniele Sicoli\affil{1}\thanks{Current address: 1306 Evansdale Drive, Morgantown, WV 26506}, Sriram Narayanan\affil{1}, and Piyush M. Mehta\affil{1}}

\affiliation{1}{Department of Mechanical, Materials and Aerospace Engineering, West Virginia University, Morgantown, WV, USA. Email: ds00146@mix.wvu.edu}

\correspondingauthor{Piyush M. Mehta}{piyush.mehta@mail.wvu.edu}

\begin{keypoints}
\item Overcrowding of LEO requires the development of new SSA tools including accurate uncertainty aware surrogate density models
\item An autoregressive SINDy (SINDy-AR) nonlinear regression framework emulates TIEGCM and other models 
\item We demonstrate stability and reliability of the framework in operations and for space situational awareness purposes
\end{keypoints}

\begin{abstract}
The ever-growing number of satellites and resident space objects in low Earth orbit keeps increasing the risk of collision between these objects. To mitigate this risk, the development of new space situational awareness tools and a new generation of probabilistic emulators of the thermosphere is increasingly felt by the community of operators and scientists. 
We previously built a Reduced Order Probabilistic Emulator based on 3 steps: dimensionality reduction, dynamic modeling, and uncertainty quantification. Here we further develop the dynamic modeling side within ROPE using a class of stable and autoregressive nonlinear models inspired by SINDy, a framework for the identification of the nonlinear dynamics. We use this framework, which we call SINDy-AR, to emulate the thermospheric density simulated by TIEGCM. This is also the first time a SINDy-AR framework is applied to such a high dimensional system as the thermospheric emulation because of its complexity. The dynamical emulators are selected on stability and accuracy criteria, so that they can be relied on in operations and space situational awareness applications. Building surrogate models with SINDy-AR can show dynamical differences between the learned emulations. 
The state-dependent structure of the TIEGCM emulation shows compatibility with the presence of terdiurnal, 27-day linked to solar rotation, semiannual, and annual tidal waves.
The execution speedup of the whole SINDy-AR pipeline with respect to TIEGCM 2.0 is about 90,000 times, which makes the new framework capable of being deployed on board a satellite given the low computational constraints it would be subjected to.
\end{abstract}

\section*{Plain Language Summary}
The increase in satellite population in low Earth orbit increases the risk of collisions between objects, thus requiring the development of the next generation of probabilistic emulators and space situational awareness tools to mitigate this risk. The largest source of uncertainty in the determination of a trajectory below altitudes of 1000 km is thermospheric density, which is largely affected by the Sun's activity. Thus, geomagnetic storms and the nonlinear activity of the Sun increase the uncertainty of risks related to variations in thermospheric density. 
We create a surrogate model of TIEGCM using an autoregressive nonlinear dynamics identification framework and perform stability analysis to select the models that form the probabilistic emulator. The way the density is generated allows uncertainty to be a natural byproduct of the framework.

\section{Introduction}
\label{sec:introduction}
\begin{figure}
\vspace{-10pt}
\centering
\includegraphics[width=1\linewidth]{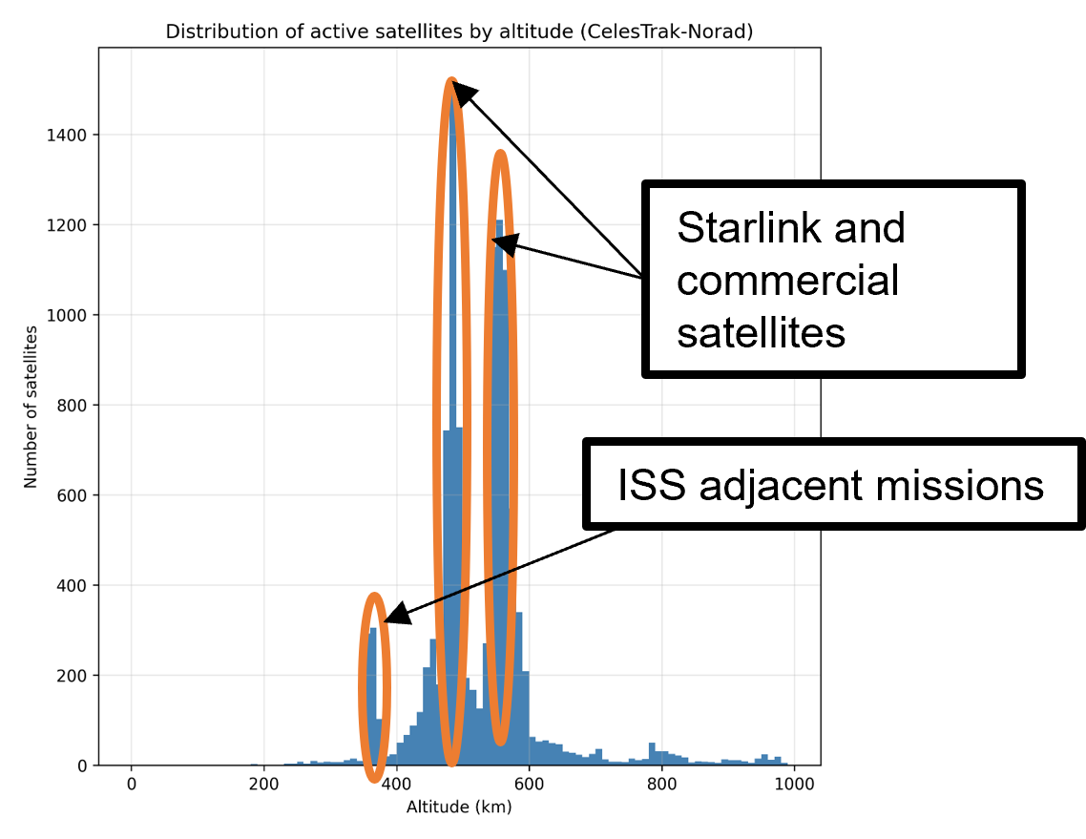}
\caption{Distribution of LEO population (credit: Celestrak). The histogram represents the number of satellites orbiting at a particular altitude. For instance, we observe the missions related to the International Space Station (ISS) slightly below 400 km of altitude, and the other Starlink and commercial satellites at 500-600 km of altitude.}
\label{fig:leo_population}
\vspace{-10pt}
\end{figure}

The number of satellites in Low Earth Orbit (LEO) has been increasing exponentially in recent years as shown in Figure \ref{fig:leo_population}, and this trend is expected to continue. The global space economy is estimated to be worth about 1.8 trillion dollars by 2035 \cite{mckinsey2023space}, which is roughly three times today's value over the next 10 years. This indicates that the LEO space economy is growing massively, as are the associated risks and opportunities, pushing forward research and innovation in the field.

The growing number of Resident Space Objects (RSOs) makes the probability of undesired events, such as collisions and maneuvers required to avoid them, larger every month that passes. Satellite overpopulation requires the development of space situational awareness techniques to mitigate the risk of collision between RSOs \cite{geostorm_threats}. We focus on the design of the next generation probabilistic emulators and of all the elements required to make it work. We built the first version of a Reduced Order Probabilistic Emulator (ROPE), \cite{ml_rope, rope_dim_red, rope_dim_red_2} based on different kinds of models.

This work develops a set of Reduced Order Models (ROMs) to emulate the thermosphere. The subclass of dynamic models of the ROPE framework introduced in this paper is composed by a set of nonlinear regression models called SINDy-AR. Model uncertainty is a natural byproduct of this framework unlike what happens with physics-based and empirical models. The goal is to dynamically emulate the global state of the full altitude TIEGCM 2.0 (Thermosphere Ionosphere Electrodynamics General Circulation Model), reaching an altitude of 980 km. This work builds upon a previous version of the framework \cite{sindy0}, called SINDyc, that was used to learn TIEGCM up to altitudes of 450 km. The full LEO coverage allows the framework to be used also by higher-altitude constellations, not just sub-450 km operations.

The Reduced Order Modeling (ROM) structure maps the higher number of degrees of freedom of the original thermosphere to a lower dimensional latent space. Once in the latent space, an AutoRegressive - Sparse Identification of Nonlinear Dynamics with control framework (SINDy-AR) \cite{sindy_pnas,sindy_kaiser,sindy_science_direct} is used to propagate the global state. The framework defines hyperparameter-dependent models, and the best fit models are selected and listed using the training dataset (see Tables \ref{tab:mape_year_3d_pca_tiegcm} and \ref{tab:mape_year_5d_pca_tiegcm}).

\begin{figure}[h]
\centering
\includegraphics[width=1\textwidth]{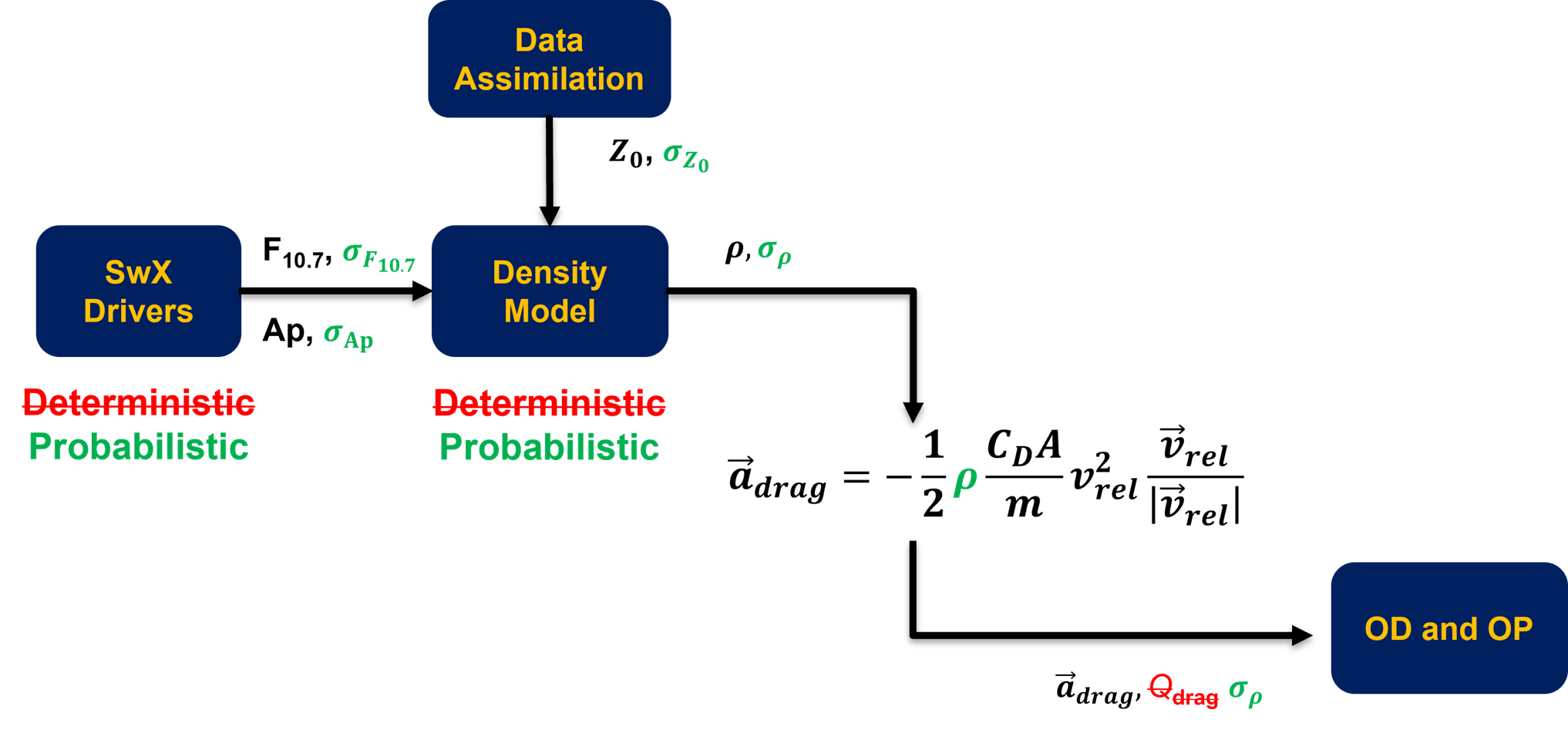}
\caption{The work the lab is dealing with to achieve the group's goals. From space weather drivers to orbit determination and propagation, going through data assimilation and dynamic modeling, which is the topic of this work. One of the achievements of our contributions is the development of a probabilistic framework.}
\label{fig:framework}
\end{figure}

The ROPE-SINDy-AR emulator has 3 parts: latent space compression, dynamic modeling, and uncertainty quantification. 
The neutral density emulation fits within other efforts, schematically represented in Figure ~\ref{fig:framework}. There is ongoing effort toward probabilistic drivers' prediction \cite{probabilistic_drivers_2023,probabilistic_drivers_2024} as well as on data assimilation \cite{rope_uncertainty,discrete_to_continuous_mehta,da_pca_sriram}. The same SINDy-AR framework builds upon other dynamic modeling and uncertainty quantification works \cite{nl_dmd,probabilistic_drivers_2023,probabilistic_drivers_2024,uncertainty_techniques_for_sw_modeling,rope_dim_red,rope_dim_red_2,ml_rope,msis_uq,sindy0}.

Thermospheric density emulation is currently offered through 4 main alternatives:

\begin{itemize}
    \item [1] Empirical models like JB2008 or NRL-MSIS 2.1 \cite{msis_nrl,msis_uq}
    \item [2] Physics-based models like the full altitude TIEGCM 2.0 \cite{tie_gcm,rope_dim_red} and WAM (Whole Atmospheric Model)
    \item [3] Data Driven probabilistic emulators like ROPE \cite{nl_dmd,rope_dim_red} and its underlying models such as the ones described here and available at \cite{sicoli_interpretable_models}, in \cite{sindy0}, or in \cite{rope_dim_red_2}.
    \item [4] The HASDM dataset, which is not publicly available in real time.
\end{itemize}

Forecast fidelity is an intrinsic property of the original model and is determined by the physical principles on which the model is based. Physics-based models generally provide the highest forecast fidelity, but at a high computational cost. Based on our experience, which also depends case by case and on HPC computational availability, TIEGCM requires approximately two weeks of computation to simulate one year of the thermosphere. The $5.5\cdot 10^5$ times speedup is based on this comparison (see Section \ref{sec:prop_times}), which does not take into account that our personal experience of the TIEGCM 2.0 run requires a HPC, while we used SINDy and measured its performances on a local machine. By running both TIEGCM and SINDy on the same machine, the speedup would be much greater than $5.5\cdot 10^5$. This substantial difference in computational cost makes uncertainty quantification challenging within the framework of physics-based models.
Empirical models like MSIS 2.1 and JB2008 are much faster than physics-based models but lack physical foundations and uncertainty quantification does not come out naturally with them.

SINDy-AR belongs to the class of data-driven emulators, learning a ``relation" between a set of input features (space weather drivers) and a global density state, and representing this relation dynamically. The dynamical relation also has physical meaning when the density state is the output of a physics-based model.

The accuracy with which the density state is known (now-cast) can be improved by applying data assimilation, see for instance \cite{da_pca_sriram}, showing data assimilation applied to \cite{sicoli_interpretable_models}.

HASDM is a dynamically-calibrated density database operated by the U.S. Space Force and distributed by Space Environment Technologies, representing the most accurate now-cast of the thermospheric density available.
HASDM is restricted from public operational use.

The SINDy-AR framework can learn the relationship between a set of drivers and the corresponding global density state, generate uncertainty through its multiple density outputs, and run with representative execution times reported in Section \ref{sec:prop_times}.

The pipeline is elastic enough that it can be used to learn a general dynamical relation $\left(F_{10}, k_p\right) \rightarrow \rho$.

\begin{figure}[H]
\centering
\includegraphics[width=1\textwidth]{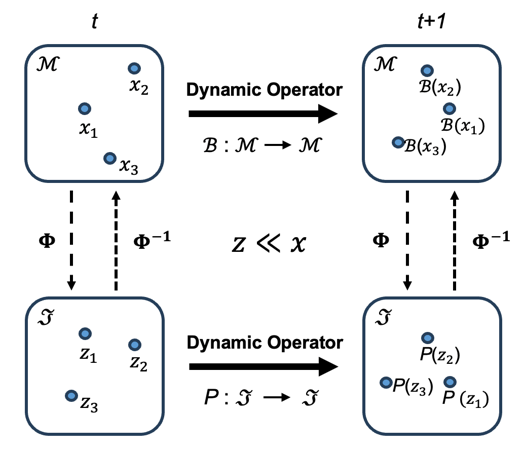}
\caption{The 3-step ROPE framework. Dimensionality reduction is done after applying a transformation map $\phi$. The reduced order (latent) space vector $Z\in \mathbb{R}^r\subset \mathbb{R}^n$ is such that $r\ll n$. The dynamic modeling step is represented by the operator P in the latent space and by the operator B in measurement space.}
\label{fig:rope}
\end{figure}

We develop the basics for a Reduced Order Modeling framework (ROM), as represented in Figure \ref{fig:rope}, which is also the basis of the ROPE emulator \cite{rope_dim_red,rope_dim_red_2}. As previously anticipated, the ROM algorithm is based on three steps.

\begin{itemize}
    \item [1] Dimensionality reduction.
    \item [2] Dynamic modeling in a reduced (or latent) state space.
    \item [3] Uncertainty quantification through ensemble modeling.
\end{itemize}

Dimensionality reduction can be linear (PCA) or nonlinear (autoencoders). Here we use PCA. The dynamic modeling step uses an autoregressive framework and employs the STLSQ (Sequentially Thresholded Least Squares) procedure to improve accuracy. A stability analysis using the spectral radius of the companion matrix and Attracting Trapping Regions (ATR) as described in \cite{atr_stability} is implemented to select the models. 
Some of the models developed in this study are contained in the online archive \cite{sicoli_interpretable_models}. \cite{da_pca_sriram} developed a data assimilation framework upon these models and the results are contained in the dataset \cite{da_dataset_pca_sriram}.

The time required to emulate the global thermospheric density starting from a particular set of initial conditions is called propagation time. Operational applications range from space situational awareness services to actionable intelligence for operators. To be ready for operational applications, the framework must satisfy a strict set of criteria regarding execution times, emulation accuracy, and ability to deliver uncertainty quantification.


We want to contribute to the community through this work with the introduction of the SINDy-AR framework, their models selection based on stability and response, thus setting the basis to prove that SINDy-AR can be reliable in operations.

The paper begins with the Introduction (Section~\ref{sec:introduction}), followed by the Methodology (Section~\ref{sec:methodology}), which describes the techniques, models, dynamical analysis and comparisons, and evaluation metrics used throughout this work.
Section~\ref{sec:results} presents the global results of the nonlinear SINDy-AR framework, including the evaluations based on global performance metrics and analysis during geomagnetic storms.
Section \ref{sec:prop_times} discusses the computational costs and how to deliver actionable intelligence to operators. Section number \ref{sec:conclusion}, which is the last one, discusses conclusions and future work.

\section{Methodology}
\label{sec:methodology}
This section describes the methodological implementations of the current work and the differences and evolutions with the methods used in \cite{rope_dim_red_2} and \cite{sindy0}. 

\subsection{Dataset}
\label{subsec:data}

\begin{sidewaysfigure}
    \centering
    \includegraphics[width=\textwidth]{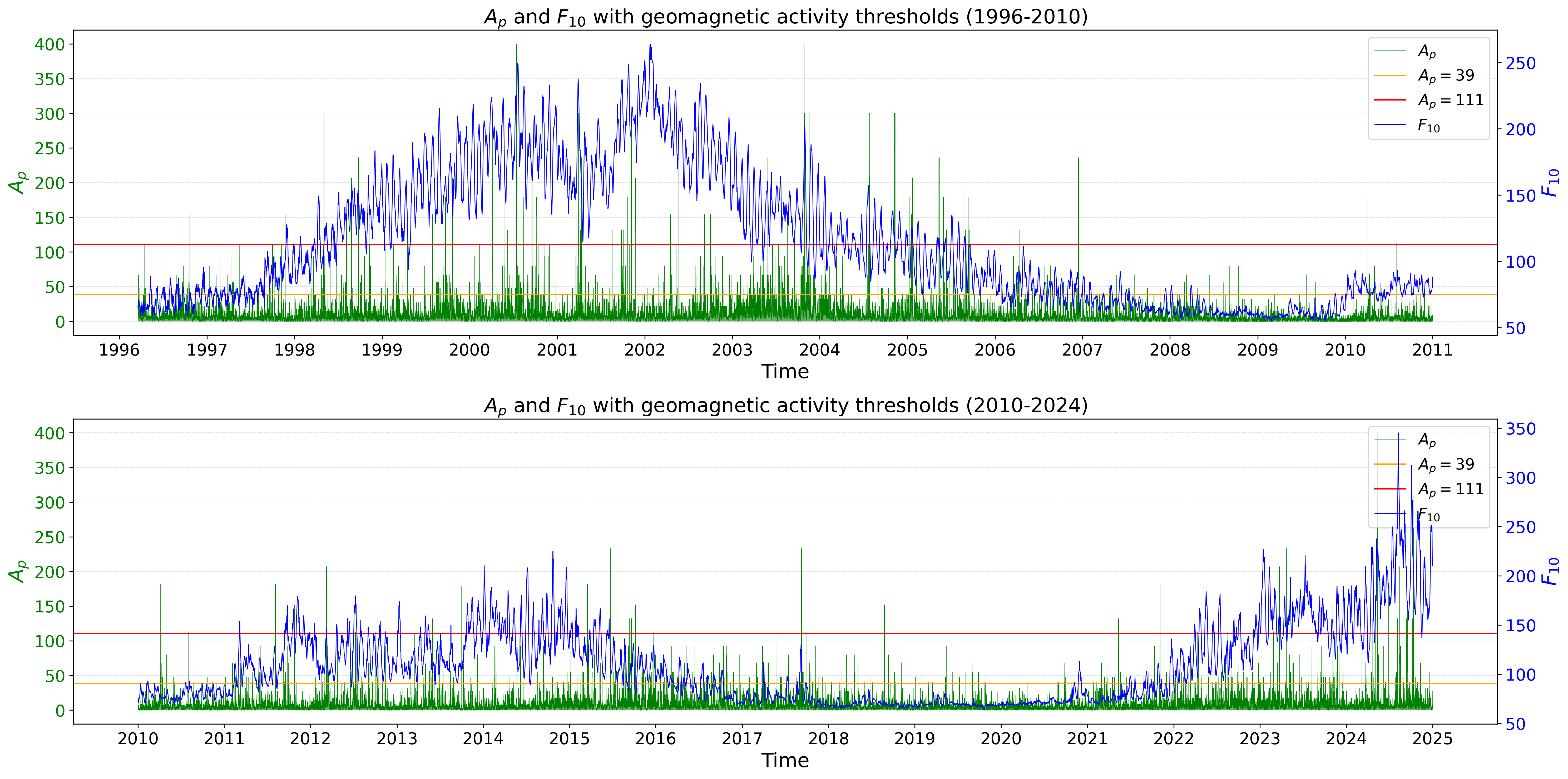}
    \caption{Drivers across the TIEGCM dataset, embracing the $23^{\text{rd}}$, $24^{\text{th}}$, and half of the $25^{\text{th}}$ solar cycle. The TIEGCM emulation uses the interval from 1996 to 2001 as test dataset, and 2002 to 2009 as train dataset. The main fundamental drivers are $F_{10}$ and $A_p$.}
    \label{fig:drivers}
\end{sidewaysfigure}

The TIEGCM dataset uses the same preprocessing structure that was implemented in works \cite{sindy0,rope_dim_red_2}. Relative to the previous works, the preprocessing pipeline uses a system of "fast datasets" that can host any kind of physics-based or data-assimilated model.
The TIEGCM emulator is trained on data from 2002 to 2009 and tested on the interval 1996-2001. The unusual train-test split for TIEGCM is due to the larger number of dynamical configurations contained in the 2002-2009 interval. Training across more widely spread dynamical configurations allows for greater generalization ability.

The TIEGCM dataset has been extrapolated to 980 km of altitude (see work \cite{rope_dim_red} or \cite{sindy0} for the corresponding methodology and PCA analysis results). 
The dataset is thus made to cover the whole LEO region, with the uniformly spaced grid having sizes

\begin{itemize}
    \item [1] Local Time (LT) values $\in [0, 23.66]$ hours for a total of 73 values with 0.33 hours variations.
    \item [2] Latitude (Lat) levels $\in [-87.5^\circ, 87.5^\circ]$ for a total of 36 values with 5 deg variations.
    \item [3] Altitude (Alt) levels $\in [0, 980]$ km for a total of 45 levels with 20 km variations.
\end{itemize}

Figure \ref{fig:drivers} shows the two basic drivers that are used to build the emulations, $F_{10}$ and $A_p$. The drivers are all public, and their sources are specified in \ref{sec:data_availability}. $F_{10}$ is the solar radio flux measured at a wavelength of 10.7 cm and indicates how the dynamics of the thermosphere expands and contracts in its linear regime. $A_p$ is the linearization of $k_p$, the planetary geomagnetic index, and can be used as a proxy to describe the nonlinear behavior of the thermosphere. They are both global planetary indices.

\subsection{Data Preparation}

To perform the emulation, solar drivers have been combined with temporal drivers, autoregressive solar drivers, autoregressive latent space variables, and finally mixed in a nonlinear regression framework. The nonlinearity of the framework is outlined by considering as input features all the nonlinear combinations of the basic set of variables up to a pre-defined polynomial order. The current investigation uses second order polynomial basis functions.    

The main temporal variables of the basic set of inputs are built using the day of the year (DOY) and hour of day (hrs), which can be summarized in the vector

\begin{align}
\label{eq:time_vars}
    \begin{cases}
        t_1 = \cos\left(\frac{2\pi\cdot \text{hrs}}{24}\right)\\
        t_2 = \sin\left(\frac{2\pi\cdot \text{hrs}}{24}\right)\\
        t_3 = \cos\left(\frac{2\pi\cdot \text{DOY}}{365.25}\right)\\
        t_4 = \sin\left(\frac{2\pi\cdot \text{DOY}}{365.25}\right)
    \end{cases}.
\end{align}

Aggregating the variables in Eq. \ref{eq:time_vars} with $F_{10}$ and $A_p$, the vector of external drivers is defined as

\begin{align}
\label{eq:drivers}
    U =
        \begin{bmatrix}
            t_1, t_2, t_3, t_4, F_{10}, A_p
        \end{bmatrix}^T
\end{align}

\subsection{Reduced Order Modeling}

The measurement space representation of the thermospheric density is

\begin{align}
    \rho_{ijk}, \text{ where } i \in [0, 71] \text{ LT }\,\ j\in [0, 35] \text{ Lat } \, \ k\in [0, 44] \text{ Alt } 
\end{align}

with a total size of 116,640 grid points for the extrapolated TIEGCM.

Modeling the TIEGCM 2.0 system in measurement space would require 116,640 dynamical equations, one equation for each of the degrees of freedom of the system in measurement space, thus posing computational challenges when these models need to be used for tasks requiring quick responses. 
Running the original TIEGCM model is 90,000 times slower than propagating the emulated dynamics with the SINDy-AR system. The learned dynamics can in fact run locally across many years at this low computational cost. The result is a high accuracy emulation of the underlying model with uncertainty quantification given as a natural byproduct of the framework itself.

There are 2 different reduction methods being used as the connection to the latent space here: Principal Component Analysis (PCA) and  Convolutional Orthogonal Autoencoders (COAE).
The first method links the original dynamics to a 10 dimensional latent space through PCA, which is a linear reduction method. Within this framework, the original vector with n degrees of freedom is mapped through a rotation in the plane to a new n dimensional vector. The components or modes of this new vector are uncorrelated elements ranked from the first to the last in terms of explained variance in decreasing order. This method is functional whenever the statistical dependence between the original components is ``linear" and is acceptable within this work to a certain extent. However, when the statistical dependence between components includes nonlinear terms as well, a nonlinear reduction method becomes more useful to identify the main dynamical modes. 

With PCA, once the rotation has been applied to the original vector (or tensor) in measurement space ($\frac{kg}{m^3}$), we cut the latent space to the initial 10 components as they explain more than $99\%$ of the variability, as shown in \cite{nl_dmd}, which also reports reconstruction errors.

In the specific case of the thermospheric density, the transformation involves a log space transformation 
\begin{align}
    \log\left(\rho \right),
\end{align}
removal of the mean value from the log-density
\begin{align}
    \log\left(\rho \right) - \mu_{\text{log}\left(\rho\right)},
\end{align}
and finally applying the PCA rotation to the transformed vector
\begin{align}
\label{eq:pca}
    Z = U^T\left(\log\left(\rho \right) - \mu_{\text{log}\left(\rho\right)}\right).
\end{align}
Using PCA allows the interpolation of the pre-defined reconstruction matrices, thus reducing the required reconstruction time when going from the latent space to the measurement space. The 14-year PCA components are plotted in Figure \ref{fig:tiegcm_full_emulation}.

\begin{figure}[h!]
\centering
\includegraphics[width=1\textwidth]{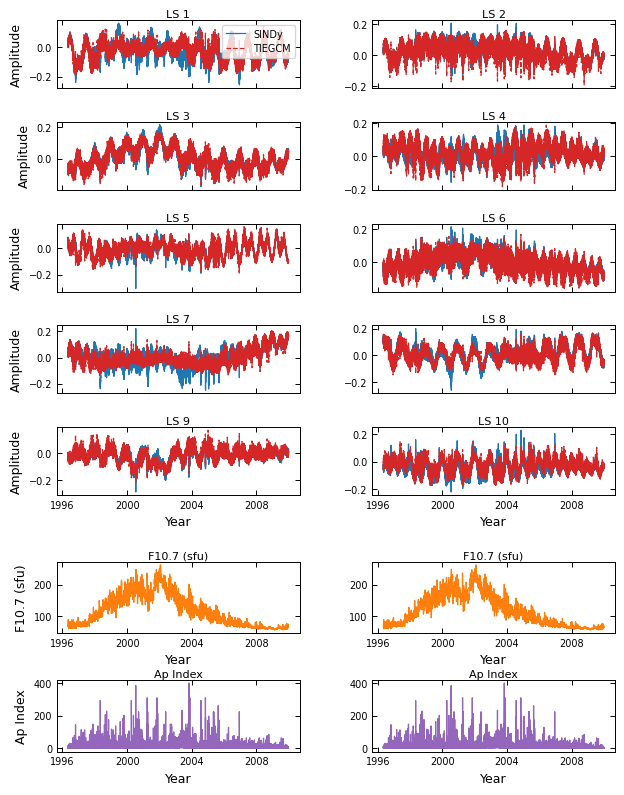}
\caption{TIEGCM emulation. Latent space components using COAE propagated using the SINDy-AR framework. The 14 year propagation completes in about 35 seconds. As opposed to PCA components, the top 1-5 components are not significantly larger than the bottom 5-10.}
\label{fig:tiegcm_full_emulation_coae}
\end{figure}

The link between measurement space and latent space in the case of orthogonal autoencoders does not assume an analytical form such as in the case of PCA. However, the 14-year latent state has been mapped as a function of time in Figure \ref{fig:tiegcm_full_emulation_coae}. Further analysis and numerical results about COAE can be found in reference \cite{rope_dim_red}.

Before training, the dependent and independent variables are standardized using

\begin{align}
    \hat{Z}_i = \frac{Z_i - \mu_{Z_i}}{\sigma_{Z_i}}
\end{align}

\subsection{DMDc}

We use DMDc, a semi-linear regression model, as a baseline to measure improvements against. The original DMDc has been outlined in \cite{nl_dmd}, and used as a baseline also in \cite{sindy0}, which contains all the details about it.
The model connects the state at the next time step to the state at the previous time step through the state coupling matrix and external drivers ($U$ as in Eq. \ref{eq:drivers}) 

\begin{align}
    X_{t+1} = A_{DMD} X_t + B_{DMD} U_t.
\end{align}

The term semi-linear depends on the augmentation of the linear variables with $F_{10}\cdot A_p$ and $A_p^2$

\begin{align}
\label{eq:dmd_drivers}
    U =
        \begin{bmatrix}
            t_1, t_2, t_3, t_4, F_{10}, A_p, F_{10}\cdot A_p, A_p^2 \\
        \end{bmatrix}^T
\end{align}

as outlined also in \cite{nl_dmd}.

\subsection{SINDy-AR}
\label{subsec:sindy}
To capture the linear and nonlinear expansion and contraction of the thermosphere we develop a nonlinear modeling framework based on SINDy \cite{sindy_kaiser}, used to identify the main features of the dynamics in the latent space.
This is the first time that this framework has been applied in such a way to a high dimensional system like the thermosphere.

The first assumption we used is that the system can be represented by a small or finite number of basis functions. However, this assumption is not always true in the real world. These basis functions can be of any kind, from polynomials through periodic functions to exponential functions. For this particular work, we are using polynomial functions. The main reason for this choice is that every sufficiently smooth function can be locally approximated by polynomials close enough to the initial point.
After fixing the kind of basis functions, we build combinations with replacement of all the input variables of the regression model. 

Another of the assumptions that we use is the sparsity of the regression matrix (see \cite{sindy_pnas}). To achieve sparsity, an STLSQ procedure is used along with ridge regression ($L^2$ regularization). Ridge regularization had already been used in \cite{sindy0}. 
The analytic formula for the ridge regression is 

\begin{equation}
    B = ZX^T(XX^T + \alpha \text{I}_{\mathbb{R}^{q \times q}})^{-1} , \ \text{where } B\in \mathbb{R}^{n\times r} \ \text{\cite{ridge_regression}}
\end{equation}

In our experiments we found that the STLSQ procedure is useful not only to achieve more sparsity of the regression matrix, but also for higher accuracy.
We implement STLSQ by ruling out all the regression coefficients below a pre-defined threshold. Once the corresponding variables have been identified, the model is retrained without those variables. The algorithm is designed such that either the framework converges within a pre-defined number of iterations, or when that number is reached, it returns the last available values.

In the experiments, we noticed that the STLSQ procedure can be used to increase the sparsity of the system at the cost of giving up some accuracy. A finite set of pre-determined basis functions may reach higher accuracies and be unstable. This motivates the following Section \ref{sec:stability} on stability analysis. This work only uses the models that are globally stable or present an Attracting Trapping Region (ATR) with finite radius \cite{atr_stability}.

The final SINDy-AR model can be represented as
\begin{equation}
    \label{eq:sindy_eq}
    Z_{k+1} = A \cdot Z_{k} + \sum_{j = 1}^{n_{AR}} A_{\text{AR}, k-j} \cdot Z_{k-j} + B U_k + \sum_{j = 1}^{n_{AR}} B_{\text{AR}, k-j} \cdot U_{k-j} + \Xi \Theta(Z_k, U_k)\ ,
\end{equation}

where $A$ is the state coupling matrix, which can also be viewed as the zeroth order set of coefficients of the AR model $A_{\text{AR}, 0}$. $Z_k$ is the latent state vector at time $k$, $A_{\text{AR}, k-j}$ is the AR state coupling matrix j time units in the past, $U_k$ is the vector of external drivers at time $k$, $\Theta$ represents the set of combinations with replacement of $Z_k$ and $U_k$, and $\Xi$ is the set of nonlinear regression coefficients.
The external drivers $U_k$ are the solar drivers as described in Section \ref{subsec:data}.

\begin{figure}[H]
\centering
\includegraphics[width=1\textwidth]{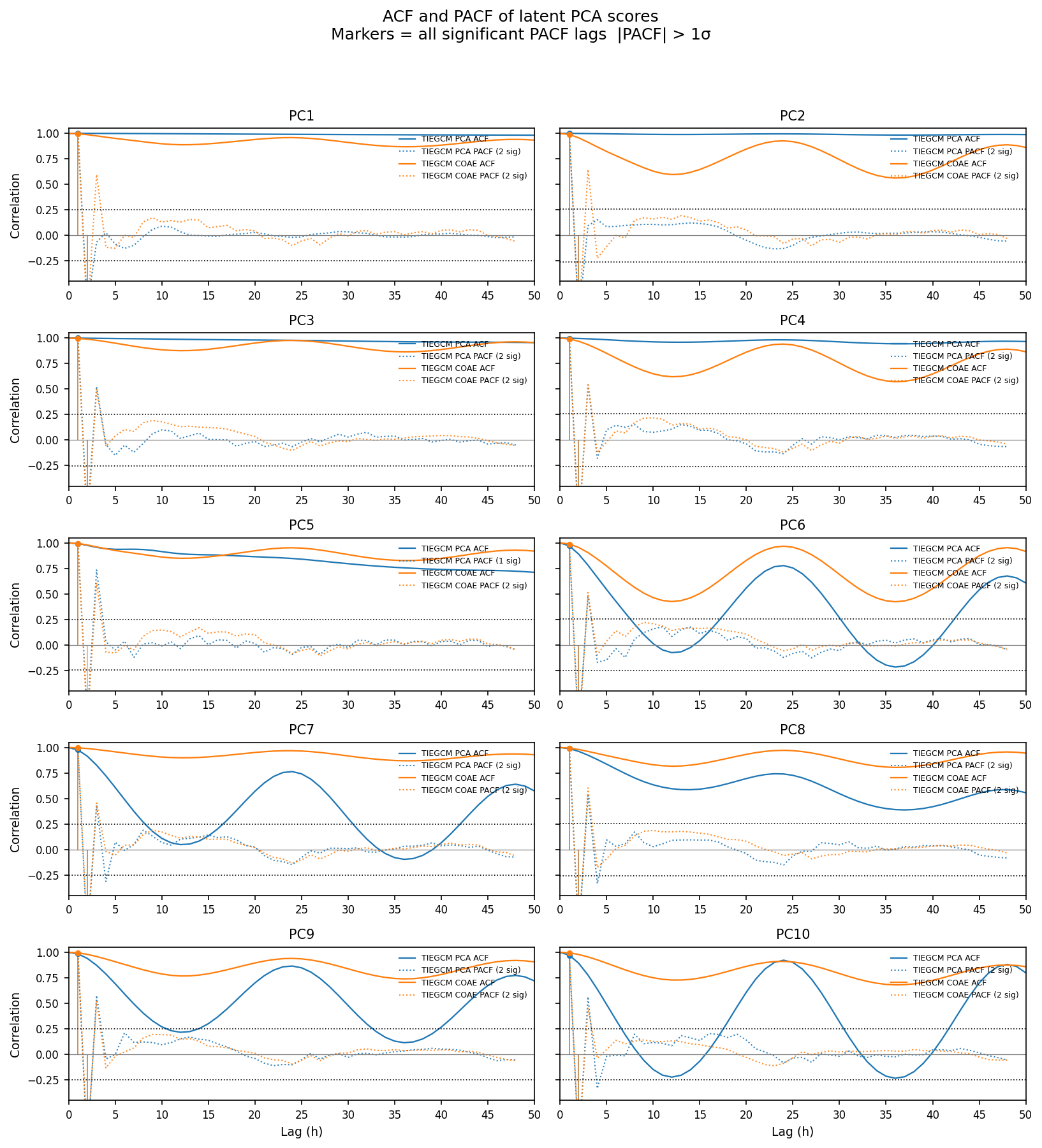}
\caption{ACF and PACF plots for the 10 latent space components of the learned TIEGCM. Blue lines represent PCA components, orange lines COAE components.}
\label{fig:acf_pacf}
\end{figure}

The steps of the autoregressive model, $n_{AR}$, is determined using Autocorrelation Function (ACF) and Partial Autocorrelation Function (PACF) analysis shown in Figure \ref{fig:acf_pacf}. The smoothness of the ACF plot demonstrates that an autoregressive modeling framework suits well the dataset. The order of the autoregressive model is determined by selecting the statistically significant peak of the PACF plot with lowest magnitude, which falls exactly 5 hours in the past.

\subsection{STLSQ-SINDy-AR, Dynamic and Stability Analysis}
\label{sec:stability}

The main contributions obtained with respect to \cite{sindy0} and the SINDy models used in \cite{rope_dim_red_2} can be summarized as

\begin{itemize}
    \item [1] The addition of autoregressive components for drivers and latent space states.
    \item [2] Fully discrete time propagation framework with adjustable time step.
    \item [3] Combinations with replacement among all input variables and not only a subset of the latent state space as in the original SINDy version \cite{sindy0}.
    \item [4] An STLSQ iterative framework improving the emulation accuracy and increasing the sparsity of the system.  
    \item [5] Stability analysis using the Attracting Trapping Region approach as described in Paper \cite{atr_stability}.
\end{itemize}

For Item 1, the experiments conducted suggested setting the autoregressive parameter to 5 hours. Selecting a memory of 5 hours in the past improved accuracy over selecting a memory of 3 hours. However, no further improvement has been observed by selecting further past memory. This empirical choice lines up with the basic lesson of time-delay embedding \cite{narayanan2025predictive}: a periodic signal needs roughly twice as many delays as the number of distinct frequencies it carries. The periodic content that dominates the emulation error is carried by only a few frequencies, so a 5-hour memory already supplies enough delays to span them, and adding more brings no measurable accuracy gain. As discussed in Section \ref{sec:spectral}, this same limit is what confines the recoverable oscillatory spectrum to a handful of modes.

Item 2 highlights an important update with respect to the SINDy framework developed in \cite{sindy0}. Noticing that discrete time step can be changed (see for instance \cite{ogata1995discrete}) pushed us to experiment in focusing in a fully discrete time framework. The way time step can be changed is described in Section \ref{sec:prop_step}.

Item 3 highlights a huge methodological improvement with respect to \cite{sindy0}, where only the drivers and the component of the latent state that is most correlated to the variations of $A_p$ was chosen.

Item 4 is required to limit the computational complexity of the models and to increase the accuracy. Sequentially Thresholded Least Squares (STLSQ) uses a pre-defined threshold to iteratively cut the regression coefficients below that threshold. This is done iteratively to rule out variables that provide useless or redundant information.

Item 5 highlights different classes of models, each one of which requires a specific kind of analysis. Equation \ref{eq:sindy_eq} can be unwrapped as

\begin{align}
    \label{eq:sindy_eq_unwrapped}
    Z_{k+1} &= A \cdot Z_{k} + \sum_{j = 1}^{n_{AR}} A_{\text{AR}, j} \cdot Z_{k-j} + B U_k + \sum_{j = 1}^{n_{AR}} B_{\text{AR}, j} \cdot U_{k-j} + \Xi \Theta(Z_k, U_k) = \\
    Z_{k+1} &= A \cdot Z_{k} + \sum_{j = 1}^{n_{AR}} A_{\text{AR}, j} \cdot Z_{k-j} + B U_k + \sum_{j = 1}^{n_{AR}} B_{\text{AR}, j} \cdot U_{k-j} + \\ 
    &+ H(Z_k\otimes Z_k) + H(Z_k\otimes U_k) + H(U_k\otimes U_k).
\end{align}
Equation \ref{eq:sindy_eq_unwrapped} can be represented considering matrix $A$ as the zeroth order autoregressive matrix, such that
\begin{align}
    \label{eq:sindy_eq_all_ar}
    Z_{k+1} &= \sum_{j = 0}^{n_{AR}} A_{\text{AR}, j} \cdot Z_{k-j} + B U_k + \sum_{j = 1}^{n_{AR}} B_{\text{AR}, j} \cdot U_{k-j} + \Xi \Theta(Z_k, U_k) = \\
    Z_{k+1} &= \sum_{j = 0}^{n_{AR}} A_{\text{AR}, j} \cdot Z_{k-j} + B U_k + \sum_{j = 1}^{n_{AR}} B_{\text{AR}, j} \cdot U_{k-j} + \\ 
    &+ H(Z_k\otimes Z_k) + H(Z_k\otimes U_k) + H(U_k\otimes U_k),
\end{align}
which is a useful representation for the construction of the companion matrix
\begin{align}
\label{eq:companion_matrix}
    \mathcal{C} = 
        \begin{pmatrix}
        A_{\text{AR},0} & A_{\text{AR},1} & A_{\text{AR},2} & \cdots & A_{\text{AR},n_{AR}-1} & A_{\text{AR},n_{AR}} \\
        I               & 0              & 0              & \cdots & 0                      & 0                    \\
        0               & I              & 0              & \cdots & 0                      & 0                    \\
        0               & 0              & I              & \cdots & 0                      & 0                    \\
        \vdots          & \vdots         & \vdots         & \ddots & \vdots                 & \vdots               \\
        0               & 0              & 0              & \cdots & I                      & 0
        \end{pmatrix}
    \in \mathbb{R}^{n_{AR} \cdot d \times n_{AR} \cdot d}.
\end{align}

Depending on the model class, stability can be selected or constrained. In the case of models that can be represented in the form 
\begin{align}
    \label{eq:sindy_eq_uu}
    Z_{k+1} &= \sum_{j = 0}^{n_{AR}} A_{\text{AR}, j} \cdot Z_{k-j} + B U_k + \sum_{j = 1}^{n_{AR}} B_{\text{AR}, j} \cdot U_{k-j} + \\ 
    &+ H(U_k\otimes U_k),
\end{align}
the strategy consists in calculating the nonlinear regression coefficients and the spectral radius associated with the companion matrix of the system.

A second class of regression models can be represented in the form
\begin{align}
    \label{eq:sindy_eq_uz0}
    Z_{k+1} &= \sum_{j = 0}^{n_{AR}} A_{\text{AR}, j} \cdot Z_{k-j} + B U_k + \sum_{j = 1}^{n_{AR}} B_{\text{AR}, j} \cdot U_{k-j} + \\ 
    &+ H(U_k\otimes Z_k).
\end{align}
The stability of this kind of model depends on the spectral radius of the Jacobian of the right hand side with respect to $Z_k$,
\begin{align}
    \label{eq:sindy_jac}
    \frac{\partial Z_{k+1}}{\partial Z_k} &= A_{\text{AR}, 0} +  H(U_k\otimes \text{Id}).
\end{align}
which depends on $F_{10}$, $A_p$, and on time $k$. Stability can be then achieved ``on average" by requiring that the spectral radius of the Jacobian be less than one across the training period. Moreover, one should ``trust" this training evaluation across the test sample. This does not guarantee intrinsic stability of the system since the spectral radius of the Jacobian can exceed unity during dynamical configurations not considered in training (storm conditions or different solar cycles). A scalable system that can be used in operations needs intrinsic stability, which is why we discard this subclass of models.

A third subclass of nonlinear regression models is given in the form
\begin{align}
    \label{eq:sindy_eq_uz}
    Z_{k+1} &= \sum_{j = 0}^{n_{AR}} A_{\text{AR}, j} \cdot Z_{k-j} + B U_k + \sum_{j = 1}^{n_{AR}} B_{\text{AR}, j} \cdot U_{k-j} + \\ 
    &+ H(Z_k\otimes Z_k),
\end{align}
where stability is achieved by constraining a finite Attracting Trapping Region (ATR) radius (see analysis in \cite{atr_stability}). The ATR optimization is obtained by constraining 
\begin{align}
    \label{eq:sindy_atr_constrains}
    &\sum_{l, i, j = 1}^{10} Z_{lk}H_l(Z_{ik}\otimes Z_{jk}) = 0\\
    &\rho\left(C C^T\right) < 1
\end{align}
where $\rho\left(C C^T\right)$ is the spectral norm of the companion matrix in Equation \ref{eq:companion_matrix}.
Since this class of models achieves accuracy comparable to the autoregressive models of Equation~\ref{eq:sindy_eq_uu} while requiring a more complex stability apparatus, the simpler formulation is preferred for the results presented in this work.

The spectral radius of all the companion matrices of all the reported models never exceeded unity and had an average value of 0.998 with a standard deviation of 0.001, and all the spectral radii of the ensemble associated with this basis function are lower than one. One of the consequences of the stability analysis of the spectral radius of the companion matrix and the attracting trapping region work (see Paper \cite{atr_stability}) is the ability to select intrinsically stable models. The eigenvalues analysis is performed on the emulated models, which learn the underlying dynamics through the SINDy-AR framework. Figure \ref{fig:sindy_eigenvalues_tiegcm} shows the eigenvalues for the learned model with hyperparameters $\lambda = 0.001$, $\alpha_{\text{ridge}} = 10$. The plot shows the 60 eigenvalues of the ``autoregressive state coupling matrix" which is called companion matrix, corresponding to a mixture of the modes of the 10 dimensional present state coupling matrix and to the modes associated with the five ten dimensional autoregressive matrices. 

The eigenvalues are much spread out in the complex plane for the learned TIEGCM. This concept is expanded in Section \ref{sec:emulations_analysis}. Each complex-conjugate pair of companion eigenvalues corresponds to one oscillatory frequency that the autoregressive backbone reconstructs. We make this connection precise in Section \ref{sec:spectral}.

The much larger diversity of real and complex eigenvalues signifies the much larger variety of long and short term modes that the model can propagate.

\subsection{Persistent Quantities in the Thermosphere}
\label{sec:persistent_quantities}

The thermosphere cannot be considered a conservative system in a strict sense. However, by excluding the external forcing provided by the Sun and all the interactions with the external environment, the quantities propagating for the longest time within the thermosphere are the closest concept to that of the conserved charges of the purely ``dynamical" isolated system. 

In the cases where stability was demonstrated and the accuracy requirements were met, the unforced system is described by the equations of motion

\begin{align}
    Z_{k+1} &= A \cdot Z_{k} + \sum_{j = 1}^{n_{AR}} A_{\text{AR}, k-j} \cdot Z_{k-j},
\end{align}

or

\begin{align}
    Z_{k+1} &=\sum_{j = 0}^{n_{AR}} A_{\text{AR}, k-j} \cdot Z_{k-j}.
\end{align}

In this case, stability is demonstrated by constructing the companion matrix of the associated 60 dimensional autoregressive system of order 1, shown in Equation \ref{eq:companion_matrix}. 

The system is stable when the spectral radius of the companion matrix in Equation \ref{eq:companion_matrix} is lower than 1. The subset of the dynamics that is not determined by external forcing is described by the equations

\begin{align}
    y_{k+1} = C y_k ,
\end{align}

where 

\begin{align}
    y_k = 
    \begin{pmatrix}
        Z_k \\ Z_{k-1} \\ \vdots \\ Z_{k-n_{AR}+1}
    \end{pmatrix}
    \in \mathbb{R}^{n_{AR} \cdot d}.
\end{align}

The most persistent modes of the learned dynamics, which we also call modal amplitudes, are determined using the left eigenvectors of C, such that 

\begin{align}
    \mathcal{Q}_j(k) = w_j^T y_k
\end{align}

satisfying the dynamics in such a way that 

\begin{align}
    \mathcal{Q}_j(k + 1) =  w_j^T y_{k+1} = w_j^T C y_k = \lambda_j w_j^T y_k  = \lambda_j \mathcal{Q}_j(k).
\end{align}

After m applications of the reduced dynamical operator $C$,

\begin{align}
    \mathcal{Q}_j(k + m) = \lambda_j^m \mathcal{Q}_j(k),
\end{align}

which tends to zero when

\begin{align}
    \left| \lambda_j \right|<1.
\end{align}

To investigate the behavior of modal amplitudes in measurement space, the 60 dimensional left eigenvectors are first projected onto the latent space dynamics and then transformed to measurement space.

Projection of mode j on the companion state is 

\begin{align}
    \alpha_j = w_j\cdot y_k.
\end{align}

Then the companion state can be rewritten such that

\begin{align}
    y_k = \sum_{j = 1}^{n_{\text{AR}}}\alpha_j v_j.
\end{align}

The Lyapunov energy associated with mode $j$ can be calculated by summing over time through

\begin{align}
    \label{eq:energy0}
    E_j = \sum_{n = 0}^\infty \alpha_j^2 \lambda_j^{2n}\lVert v_j \Vert_{P_0}.
\end{align}

The quantity in Equation \ref{eq:energy0} can be calculated only when $\left| \lambda_j \right|<1$ for every $j$, that is when the system is stable, which can be done with the classes of models that we are using. Calculations lead to

\begin{align}
    E_j = \frac{\alpha_j^2}{1-\left|\lambda_j\right|^2} v_j^2,
\end{align}

that can be used in the next Section \ref{sec:spectral} and for future reference.

\subsection{Spectral Interpretation and Recoverable Oscillatory Dimension}
\label{sec:spectral}

Beyond their role as a stability diagnostic, the eigenvalues of the learned companion operator are the spectral poles of the latent signal. This identification is one of the reasons that lets us say, precisely, in what sense the SINDy-AR emulator of TIEGCM recovers genuine dynamics. The propagation step is $\Delta t = 1$~hour throughout this section.

Diagonalizing the companion matrix $\mathcal{C}$ introduced above (or its unforced restriction $C$) and propagating the modal coordinates $\alpha_j$, any latent observable is a sum of damped complex exponentials,

\begin{align}
    \label{eq:modal_exp}
    z_i(k) &= \sum_{j} c_{ij}\,\lambda_j^{\,k},
    \qquad \lambda_j = |\lambda_j|\,e^{i\theta_j},\quad \theta_j = \omega_j \Delta t .
\end{align}

Taking the discrete-time Fourier transform (equivalently the one-sided $\mathcal{Z}$ transform) of Equation \ref{eq:modal_exp} and evaluating it on the unit circle $z = e^{i\omega \Delta t}$,

\begin{align}
    \label{eq:dtft_poles}
    \hat{Z}_i(\omega)
    &= \sum_{k=0}^{\infty} z_i(k)\, e^{-i\omega\Delta t\,k}
     = \sum_{k=0}^{\infty} \sum_{j} c_{ij}\, \lambda_j^{\,k}\, e^{-i\omega\Delta t\,k} \nonumber\\
    &= \sum_{j} c_{ij} \sum_{k=0}^{\infty}
       \big(\lambda_j\, e^{-i\omega\Delta t}\big)^{k}
     = \sum_{j} \frac{c_{ij}}{1 - \lambda_j\, e^{-i\omega \Delta t}},
\end{align}

where the interchange of sums and the closed form use the geometric series $\sum_{k\ge 0} r^k = (1-r)^{-1}$ with $r = \lambda_j\, e^{-i\omega\Delta t}$, which converges because the operator is stable by construction ($|r| = |\lambda_j| < 1$). This has poles exactly where $e^{i\omega \Delta t} = \lambda_j$. The eigenvalues are therefore the poles of the signal spectrum: their arguments $\theta_j = \arg(\lambda_j) = \omega_j \Delta t$ are the resonant angular frequencies, and the FFT power spectrum $|\hat{Z}_i(\omega)|^2$ exhibits a peak at $f_j = \theta_j/(2\pi\Delta t) = 1/T_j$ whose sharpness is set by $|\lambda_j|$ ($|\lambda_j|\to 1$ yields a sharp spectral line and a long half-life, while $|\lambda_j|<1$ broadens and damps it). This is the standard autoregressive, or Prony, view of linear prediction and spectral estimation \cite{stoica2005spectral,mezic2005spectral}. It is also the spectral counterpart of time-delay embedding, where the number of distinct frequencies in a periodic signal can be read off from the rank of its delay (Hankel) matrix \cite{narayanan2025predictive}; here that same count is simply the number of off-axis complex-conjugate eigenpairs.

We call a mode \emph{oscillatory} when $\mathrm{Im}\,\lambda_j \neq 0$ (a genuine nonzero frequency) and \emph{relaxational} when $\lambda_j \in \mathbb{R}_+$ ($\theta_j = 0$, a pure decay or baseline-following mode). It is worth stressing that the modulus $|\lambda_j|$ alone does not separate the two: a pole may sit arbitrarily close to the unit circle and still be purely relaxational, in which case it merely propagates driver perturbations without oscillating. What distinguishes a recovered dynamics is that a mode be \emph{off-axis} and simultaneously \emph{long-lived relative to its own period}, i.e. that its half-life $t_{1/2,j} = -\ln 2\,\Delta t/\ln|\lambda_j|$ satisfy $t_{1/2,j} \gtrsim T_j$, so that the oscillation completes at least one full cycle before damping out. This suggests two simple measures of how much dynamical content a model carries. The first is the recoverable oscillatory dimension $M_{\mathrm{osc}}$, the number of off-axis conjugate pairs above the noise floor. The second is the off-axis energy fraction

\begin{align}
    \label{eq:eta}
    \eta = \frac{\sum_{j:\,\mathrm{Im}\,\lambda_j \neq 0} E_j}{\sum_{j} E_j},
\end{align}

with the modal energy $E_j$ of Equation \ref{eq:energy0}.

Because TIEGCM integrates the governing equations, it self-generates the thermospheric tides at periods $T = 24, 12, 8, 6$~h even though the input indices carry no such lines. By Equation \ref{eq:dtft_poles}, a mode of period $T$ has angular frequency $\omega = 2\pi/T$, so its eigenvalue argument is

\begin{align}
    \label{eq:theta_tide}
    \theta = \omega \Delta t = \frac{2\pi\,\Delta t}{T},
\end{align}

which is simply the angle the mode rotates through in one propagation step. Evaluating Equation \ref{eq:theta_tide} at $\Delta t = 1$~h gives $\theta_{24} = 2\pi/24 \approx 0.262$, $\theta_{12} = 2\pi/12 \approx 0.524$, $\theta_{8} = 2\pi/8 \approx 0.785$, and $\theta_{6} = 2\pi/6 \approx 1.047$~rad. Each internally generated harmonic is therefore \emph{expected} to force a complex-conjugate eigenpair at one of these angles, sitting well off the real axis and with a half-life long compared with its own period.

What the emulator actually recovers is a subset of this ladder, and the distinction matters. As reported in Table \ref{tab:tidal_comparison}, the learned TIEGCM operator resolves the terdiurnal line as a finite-$T$ eigenpair at $8.2$~h ($\theta \approx 0.766$~rad, against the predicted $\theta_8 \approx 0.785$), together with a family of long-period modes matching the semiannual and solar-rotation bands. The table also reports the closest frequency identified as associated with annual tides, which requires a discussion addressed in \ref{sec:emulations_analysis}. The diurnal and semidiurnal harmonics, by contrast, are not resolved as finite-$T$ eigenpairs of the companion matrix (see Eq. \ref{eq:companion_matrix}) at this latent dimension and autoregressive order. Instead, the diurnal tides are directly represented by the clock built from the $t_1$ and $t_2$ variables of Equation \ref{eq:time_vars}. The semidiurnal tides are given by the clock with twice the angular frequency of $t_1$ and $t_2$, an expression that is a second order polynomial in $t_1$ and $t_2$ and is therefore perfectly recoverable by the second order basis functions being used, without requiring free dynamics. The terdiurnal tide is a third order polynomial (degree-3 harmonic) of the same clock and is not reachable by second order terms, which aligns with its appearing as a genuine eigenmode, while the diurnal and semidiurnal tides do not. Their absence is a statement about the identified reduced operator rather than about TIEGCM itself, in which the migrating tides are, of course, present; recovering them would require either a richer latent basis or a longer delay embedding, since the $5$-hour memory of Section \ref{sec:stability} supplies only enough delays to span a handful of distinct frequencies. Even so, the modes that are recovered are genuinely off-axis and survive their own period, so $M_{\mathrm{osc}}^{\text{TIEGCM}}$ is comfortably larger than zero and $\eta^{\text{TIEGCM}}$ is of order one.

It is worth being clear about what this does and does not show. It is a statement about the \emph{recoverable dynamics}: it exposes the dynamical mechanism, but on its own it does not establish superior forecasting of the real atmosphere, which would still need validation against independent observations.

\begin{figure}[h!]
  \centering
  \begin{subfigure}[b]{1\textwidth}
    \centering
    \includegraphics[width=\textwidth]{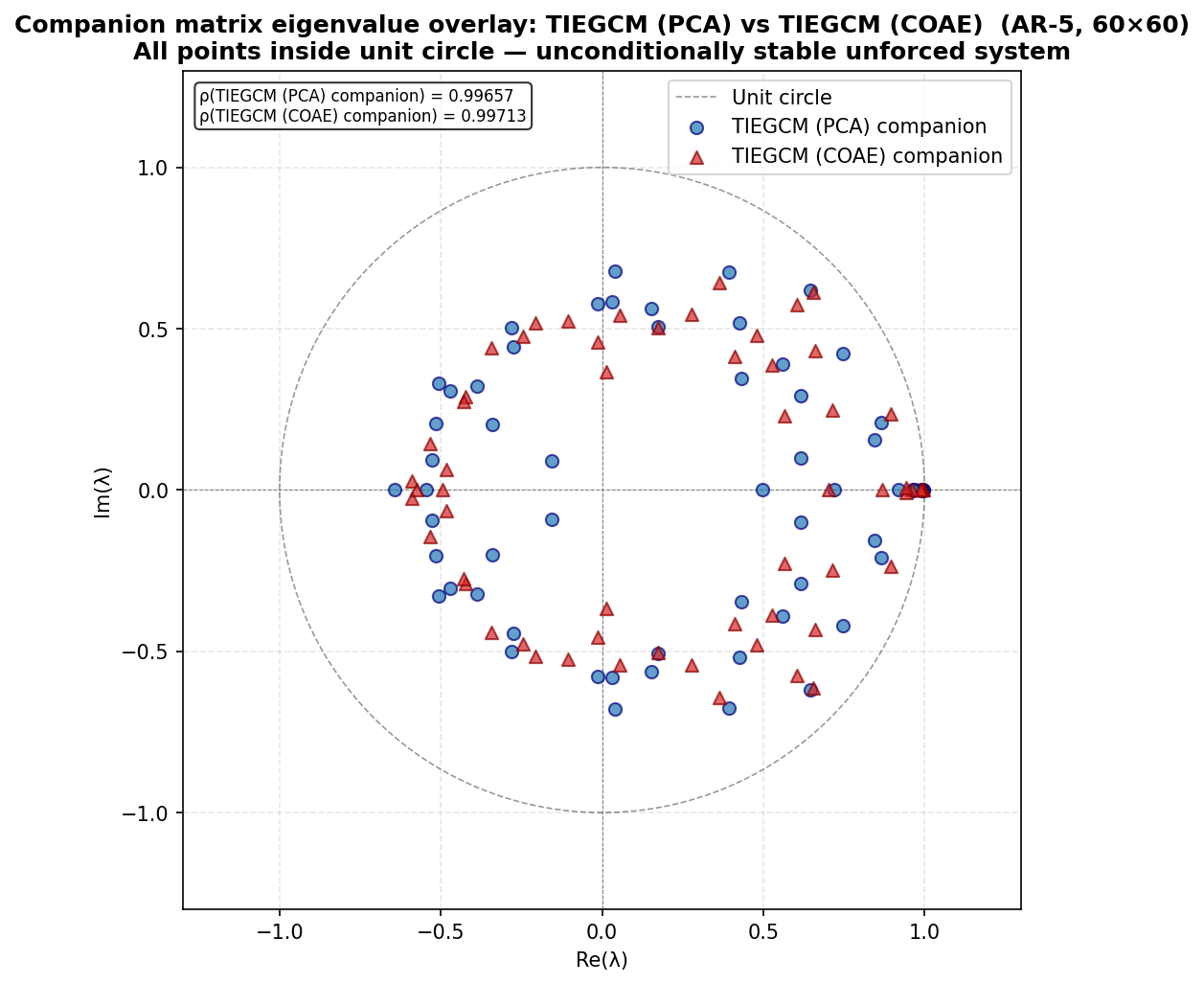}
    \caption{SINDy-AR}
    \label{fig:sindy_eigenvalues_sindy}
  \end{subfigure}
  \hfill
  \begin{subfigure}[b]{1\textwidth}
    \centering
    \includegraphics[width=\textwidth]{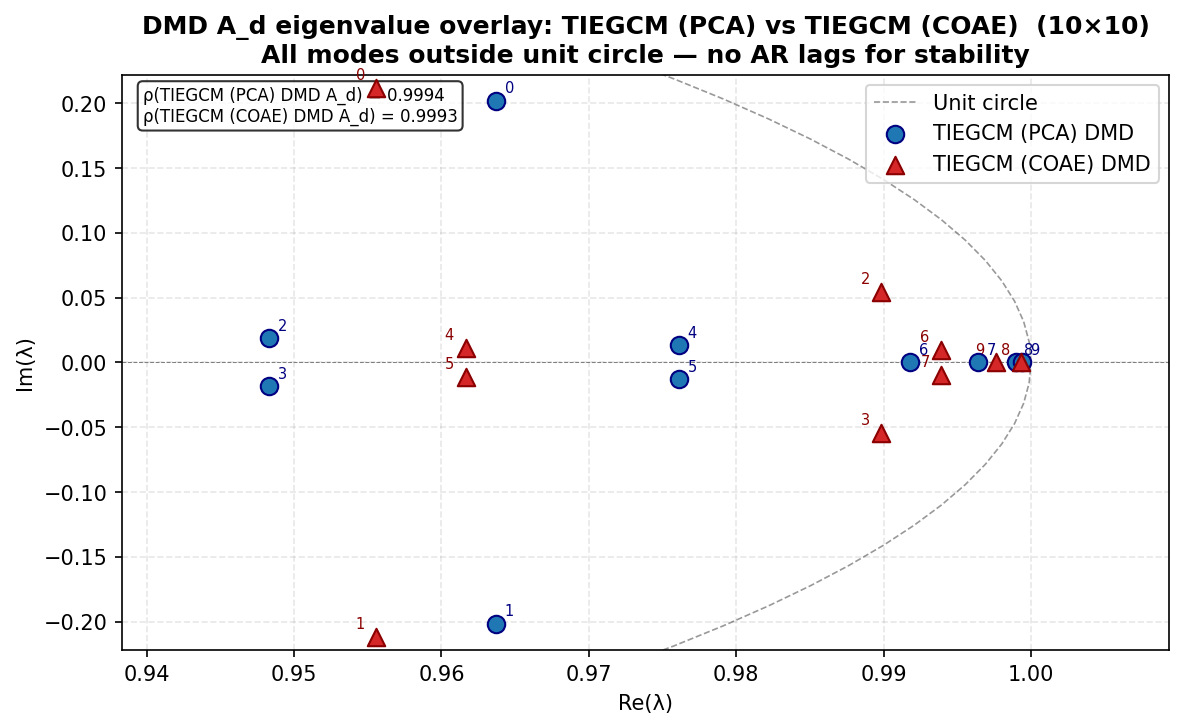}
    \caption{DMDc}
    \label{fig:sindy_eigenvalues_dmd}
  \end{subfigure}
  \caption{Model eigenvalues for full altitude TIEGCM, PCA-COAE comparison: (a) SINDy-AR, (b) DMDc.}
  \label{fig:sindy_eigenvalues_tiegcm}
\end{figure}

One last point is that no separate spectral figure is needed here. Since the FFT peak frequencies and the eigenvalue arguments have to coincide by Equation \ref{eq:dtft_poles}, the eigenvalue plot of Figure \ref{fig:sindy_eigenvalues_tiegcm} already carries the same information. The recovered TIEGCM eigenvalues sit at angles matching the tidal and long-period bands of Table \ref{tab:tidal_comparison}, with half-lives long enough for those modes to complete their period. This is just what Equations \ref{eq:dtft_poles} to \ref{eq:modal_exp} predict.

\subsection{Change of Time Step}
\label{sec:prop_step}

Calculations show \cite{ogata1995discrete} that whenever the propagation step is changed from $T_1$ to $T_2$, a system driven by the equations of motion 

\begin{align}
    Z_{k+1} = A_{T_1} \cdot Z_{k} + B_{T_1} \cdot U_k
\end{align}

valid for the $T_1$ propagation step, can be transformed to 

\begin{align}
    Z_{k+1} = A_{T_2} \cdot Z_{k} + B_{T_2} \cdot U_k
\end{align}

where

\begin{align}
    &A_{T_2} = A_{T_1}^{\frac{T_2}{T_1}}\\
    &B_{T_2} = \left(A_{T_2} - \text{Id}\right)\cdot \left(A_{T_1} - \text{Id}\right)^{-1}\cdot B_{T_1}.
\end{align}

The use of different propagation time steps serves to give end-users the elasticity to propagate the thermosphere for the required amount of time. More practically, in operations, an operator may need to know how to maneuver a satellite in the following 3 hours with a 60 second resolution. In this case, the system will provide exactly the required global state of the thermosphere propagated with the required time step. From there, as explained in Section \ref{sec:prop_times}, the operator or end-user can interpolate the global state at the desired locations in space and time.

\subsection{Evaluation Metrics}
\label{sec:metrics}
The evaluation metric used to calculate model performances is the Mean Absolute Percentage Error (MAPE(\%)).
The MAPE(\%) is calculated as
\begin{align}
    \text{MAPE(\%)} = \frac{100}{m \cdot n_{\text{LT}}\cdot n_{\text{Lat}} \cdot n_{\text{Alt}}}\sum_{t=1}^m\sum_{i=1}^{n_{\text{LT}}}\sum_{j=1}^{n_{\text{Lat}}}\sum_{k=1}^{n_{\text{Alt}}}\frac{\|\hat{\rho}_{ijkt} - \rho_{ijkt}\|}{\rho_{ijkt}},
\end{align}
with m total time steps, $n_{\text{LT}}$ number of local solar time indices, $n_{\text{Lat}}$ number of latitude indices, and $n_{\text{Alt}}$ number of altitude steps.

When comparing model performances of a model over another, we use the relative percent variation
\begin{align}
\label{eq:relative_error}
    \Delta(\%) = 100 \cdot \frac{\hat{y} - y}{y},
\end{align}
that we also define as Relative MAPE Improvement ($RMI$) when comparing MAPE results.

Another useful formula is given by the rolling standard deviation of $A_p$
\begin{align}
    \sigma(A_p) = \sqrt{\frac{1}{T-1} \sum_{t = 1}^T \left(A_p(t) - \Bar{A_p}\right)^2},
\end{align}
which can also be generalized to calculate other rolling quantities (like rolling means).


\section{Results}
\label{sec:results}
This section summarizes the results obtained with the stable SINDy-AR models trained from 2002 to 2009 and tested from 1996 to 2001 for TIEGCM.

The models are chosen to be stable by construction and are tested across all the training and test periods, which means that the results are obtained from a selection of the best models among all the stable SINDy-AR models.  

\begin{figure}[h!]
\centering
\includegraphics[width=1\textwidth]{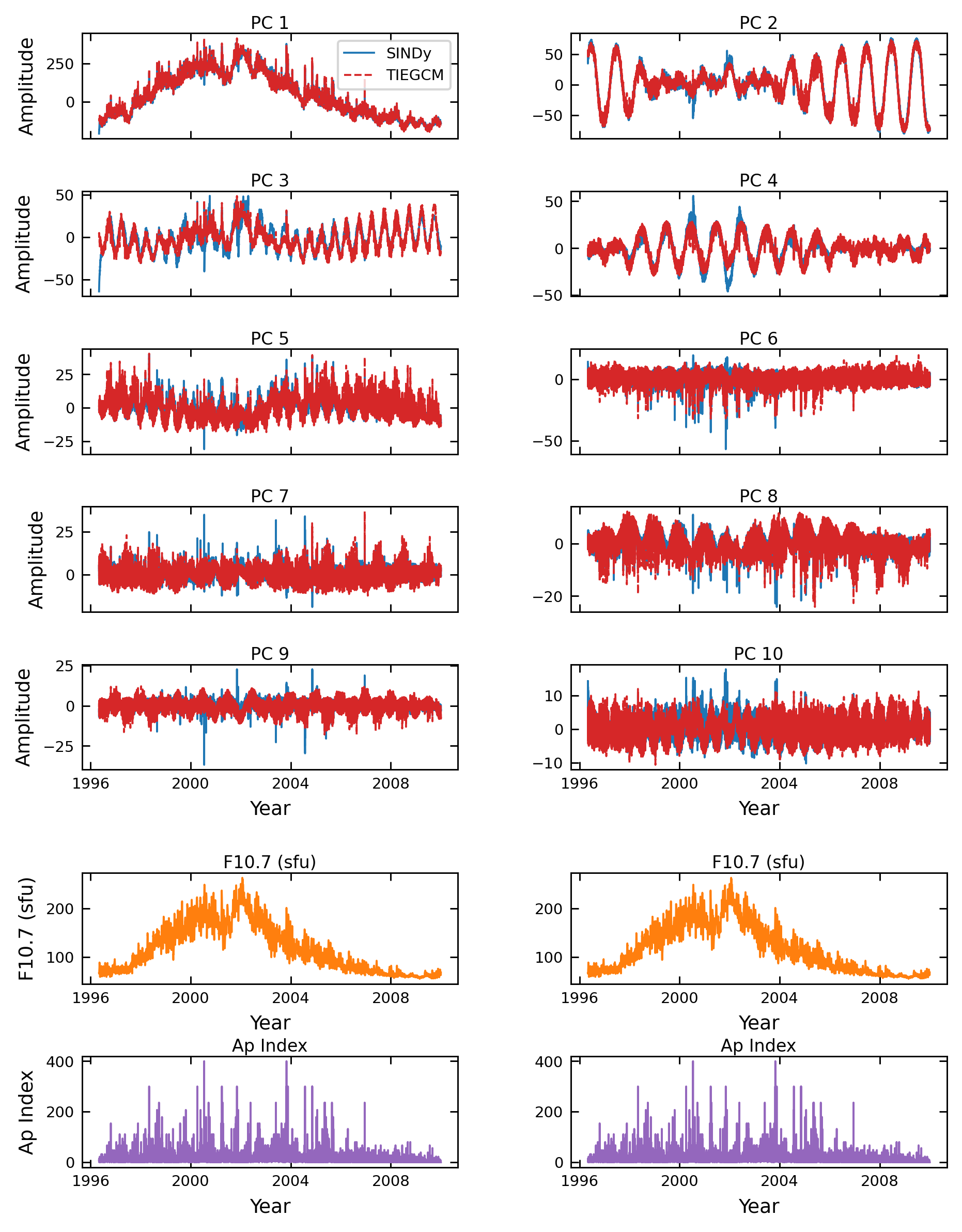}
\caption{TIEGCM emulation. Latent space components using PCA propagated using the SINDy-AR framework. The 14 year propagation in latent space completes in 31 seconds. Average correlation between model and truth $\sim 94\%$. The TIEGCM emulation shows generalization ability from train to test.}
\label{fig:tiegcm_full_emulation}
\end{figure}

An average correlation between the learned dynamics and the simulations in the latent space of 94\% demonstrates the efficiency of the framework in emulating the underlying dynamics. Dynamical emulation fidelity can be visually grasped also from Figure \ref{fig:tiegcm_full_emulation} showing a single propagation of 14 years, with no initial condition reset. This Figure represents the best among the stable SINDy-AR models (STLSQ = 0.001, ridge = 10) that are used to emulate TIEGCM.

\subsection{The Dynamical Analysis of the TIEGCM Emulation}
\label{sec:emulations_analysis}

Dynamical analysis of the surrogate models is useful to discern whether the learned dynamics is stable or unstable. Among all the dynamics, only the stable ones were retained (see Section \ref{sec:stability}) and evaluated. It was also possible to assess whether the SINDy-AR framework is capable of reproducing some of the observed physical phenomena, such as tidal waves, which are physically well characterized. In addition to allowing fast propagation of the global thermospheric state, the dynamical emulation is useful for measuring the fidelity with respect to the presence and accuracy of these physical quantities. Relatedly, Table \ref{tab:tidal_comparison} indicates how much of this kind of physical information is learned by the framework. At first glance, Table \ref{tab:tidal_comparison} shows the set of captured eigenfrequencies corresponding to the closest physical frequencies found in nature. However, the modes deserve some comment. We understand that the diurnal and semidiurnal modes are not captured by the purely intrinsic dynamics of the companion matrix, as they are exactly expressed by the application of the second order basis functions to the variables $t_1$ and $t_2$ of Equation \ref{eq:time_vars}. In fact, the diurnal tides are already in $U$ and do not need pure dynamics, since $t_1 = \cos(\omega t)$ and $t_2 = \sin(\omega t)$; and the semidiurnal tides can be obtained as second order polynomials of sines and cosines of $t_1$ and $t_2$, $\cos(2\omega t) = \cos^2(\omega t) - \sin^2(\omega t)$ and $\sin(2\omega t) = 2 \cdot \sin(\omega t)\cos(\omega t)$, and don't need pure dynamics either. A different argument must be used for the semiannual and annual tides. \label{discussion:ann_semiann}Annual and semi-annual are not pure geometry as diurnal and semi-diurnal tides since the thermosphere runs denser in January than July and has a related semiannual asymmetry that a plain day-of-year clock cannot capture \cite{Qian2009Seasonal}, so they both leave a real residual for the free dynamics to explain, which is why they show up in the table while diurnal and semidiurnal do not. The value of the annual tide is off by 33\% while the values of the other lines land in a 2-6\% confidence interval, including the semiannual period. While there are thousands of diurnal and semi-diurnal data points across 14 years that the models can use to find the right frequencies, there are roughly 14 annual observations to constrain the annual eigenvalue, and roughly twice as many semiannual cycles. A frequency this poorly sampled is going to be noisy, and that alone accounts for the gap between the clean 1.9\% semiannual match and the 33\% miss on annual.

\begin{table*}[htbp]
\centering
\caption{Comparison of physical tidal/wave periods with periods identified in the learned TIEGCM system.}
\label{tab:tidal_comparison}
\resizebox{\textwidth}{!}{%
\footnotesize
\begin{tabular}{lcc}
\toprule
\textbf{Physical Mode} & \textbf{Standard Period} & \textbf{TIEGCM Match} \\
\midrule
Terdiurnal (24h/3)             & 8.00 h  & 8.2 h (modes 8,9)      \\
Diurnal (24h)                  & 24.0 h  & ---                    \\
Semidiurnal (24h/2)            & 12.0 h  & ---                    \\
Semiannual (SAO)               & 4383 h  & 4301.7 h (modes 2,3)   \\
Annual (Unresolved: see discussion in Section \ref{discussion:ann_semiann})                        & 8766 h  & 11688.6 h (modes 0,1)  \\
\bottomrule
\end{tabular}%
}
\end{table*}

\subsection{Global Metrics Evaluation}
\label{sec:global_results}
Once the stable models have been filtered from the whole set of model guesses, the model selection is performed by MAPE evaluation as defined in Section \ref{sec:metrics}. The best models are selected by finding the models corresponding to the minimum MAPE for each STLSQ threshold by running over different hyperparameters.
Dynamic forecasting and MAPE calculation are performed on an ensemble of 3 and 5 day propagation intervals. At the beginning of every 3 or 5 day propagation interval, the initial conditions correspond to the exact density values. Dynamic forecasting is defined by using the output of the previous forecast at the next propagation step for each of the propagation steps within the 3 or 5 day propagation interval.

The reduction methods that were evaluate are PCA and COAE.
The PCA-TIEGCM emulation MAPE of the best models is about 6-7\% (see Table \ref{tab:summary_5d_tiegcm}). This quantifies the uncertainty with which the SINDy-AR framework can emulate the dynamics of the TIEGCM model.

\begin{table}[htbp]
  \centering
  \caption{TIEGCM emulation. Summary MAPE (\%), 3-day propagation window. Training: 2002--2009. Test: 1996--2001.}
  \label{tab:summary_3d_tiegcm}
  \begin{tabular}{llcccc}
    \toprule
    Model & $\lambda$ & $\alpha_{\text{ridge}}$ & Train Avg (2002--2009) & Test Avg (1996--2001) & RMI vs.\ DMDc (\%) \\
    \midrule
    F10.7 & --- & --- & 102.5 & 133.4 & --- \\
    Ap & --- & --- & 10.79 & 11.30 & --- \\
    DMDc & --- & --- & \textcolor{red}{\textbf{7.49}} & \textcolor{red}{\textbf{7.94}} & 0.00 / 0.00 \\
    \midrule
    \multirow{5}{*}{SINDy-AR} & 0.001 & 10.0 & \textcolor{blue}{\textbf{6.12}} & \textcolor{blue}{\textbf{6.18}} & 18.3 / 22.2 \\
     & 0.005 & 0.1 & 6.30 & 6.59 & 15.9 / 17.0 \\
     & 0.01 & 10.0 & 6.16 & 6.23 & 17.8 / 21.5 \\
     & 0.05 & 0.1 & 8.17 & 8.24 & $-$9.1 / $-$3.8 \\
     & 0.1 & 10.0 & 8.21 & 8.23 & $-$9.6 / $-$3.7 \\
    \bottomrule
  \end{tabular}
\end{table}

\begin{table}[htbp]
  \centering
  \caption{TIEGCM emulation. Summary MAPE (\%), 5-day propagation window. Training: 2002--2009. Test: 1996--2001.}
  \label{tab:summary_5d_tiegcm}
  \begin{tabular}{llcccc}
    \toprule
    Model & $\lambda$ & $\alpha_{\text{ridge}}$ & Train Avg (2002--2009) & Test Avg (1996--2001) & RMI vs.\ DMDc (\%) \\
    \midrule
    F10.7 & --- & --- & 102.5 & 133.4 & --- \\
    Ap & --- & --- & 10.78 & 11.29 & --- \\
    DMDc & --- & --- & \textcolor{red}{\textbf{8.51}} & \textcolor{red}{\textbf{8.99}} & 0.00 / 0.00 \\
    \midrule
    \multirow{5}{*}{SINDy-AR} & 0.001 & 10.0 & \textcolor{blue}{\textbf{6.57}} & \textcolor{blue}{\textbf{6.66}} & 22.8 / 25.9 \\
     & 0.005 & 0.1 & 6.79 & 7.06 & 20.2 / 21.5 \\
     & 0.01 & 10.0 & 6.68 & 6.78 & 21.5 / 24.6 \\
     & 0.05 & 0.1 & 9.11 & 9.05 & $-$7.1 / $-$0.7 \\
     & 0.1 & 10.0 & 9.22 & 9.35 & $-$8.3 / $-$4.0 \\
    \bottomrule
  \end{tabular}
\end{table}

The COAE-TIEGCM SINDy-AR emulation holds a surprise for us. Ideally, the COAE reduction method was designed to allow for a latent space which represented the nonlinearities of the system more effectively. However, although the COAE latent space does this job, the statistical results obtained using the COAE+SINDy-AR are slightly worse than using PCA+SINDy-AR. This is true also for the DMDc baseline. The plausible explanation is all overall improvement in the nonlinearity of the latent space dynamics.   

\begin{table}[htbp]
  \centering
  \caption{COAE-TIEGCM emulation summary, polynomial order 2 basis functions. MAPE (\%), 3-day propagation window. Training: 2002--2009. Test: 1996--2001.}
  \label{tab:summary_3d_cae_p2}
  \begin{tabular}{llcccc}
    \toprule
    Model & $\lambda$ & $\alpha_{\text{ridge}}$ & Train Avg (2002--2009) & Test Avg (1996--2001) & RMI vs.\ DMDc (\%) \\
    \midrule
    F10.7 & --- & --- & 102.5 & 133.4 & --- \\
    Ap & --- & --- & 10.81 & 11.29 & --- \\
    DMDc & --- & --- & \textcolor{red}{\textbf{9.80}} & \textcolor{red}{\textbf{10.05}} & 0.00 / 0.00 \\
    \midrule
    \multirow{3}{*}{SINDy-AR} & 0.001 & 10.0 & \textcolor{blue}{\textbf{6.62}} & \textcolor{blue}{\textbf{6.81}} & 32.4 / 32.2 \\
     & 0.005 & 0.1 & 6.97 & 6.91 & 28.9 / 31.2 \\
     & 0.01 & 10.0 & 6.65 & 6.75 & 32.1 / 32.8 \\
    \bottomrule
  \end{tabular}
\end{table}

\begin{table}[htbp]
  \centering
  \caption{COAE-TIEGCM emulation summary, polynomial order 3 basis functions. MAPE (\%), 3-day propagation window. Training: 2002--2009. Test: 1996--2001.}
  \label{tab:summary_3d_cae_p3}
  \begin{tabular}{llcccc}
    \toprule
    Model & $\lambda$ & $\alpha_{\text{ridge}}$ & Train Avg (2002--2009) & Test Avg (1996--2001) & RMI vs.\ DMDc (\%) \\
    \midrule
    F10.7 & --- & --- & 102.5 & 133.4 & --- \\
    Ap & --- & --- & 10.81 & 11.29 & --- \\
    DMDc & --- & --- & \textcolor{red}{\textbf{9.80}} & \textcolor{red}{\textbf{10.05}} & 0.00 / 0.00 \\
    \midrule
    \multirow{3}{*}{SINDy-AR} & 0.001 & 10.0 & \textcolor{blue}{\textbf{5.65}} & \textcolor{blue}{\textbf{6.28}} & 42.3 / 37.5 \\
     & 0.005 & 0.1 & 5.82 & 6.27 & 40.6 / 37.6 \\
     & 0.01 & 10.0 & 5.95 & 6.46 & 39.3 / 35.7 \\
    \bottomrule
  \end{tabular}
\end{table}

The SINDy-AR framework improves by about 14\% when training a third order basis function model (Table \ref{tab:summary_3d_cae_p3}) as opposed to a second order basis function model (Table \ref{tab:summary_3d_cae_p2}) by looking at the 3-day dynamic forecast Tables. The same relative improvement rate is lower in the test dataset. This means that increasing the nonlinearity of the fitting basis functions can improve the situation, although this happens more slowly within the test dataset.

\begin{table}[htbp]
  \centering
  \caption{COAE-TIEGCM emulation summary, polynomial order 2 basis functions. MAPE (\%), 5-day propagation window. Training: 2002--2009. Test: 1996--2001.}
  \label{tab:summary_5d_cae_p2}
  \begin{tabular}{llcccc}
    \toprule
    Model & $\lambda$ & $\alpha_{\text{ridge}}$ & Train Avg (2002--2009) & Test Avg (1996--2001) & RMI vs.\ DMDc (\%) \\
    \midrule
    F10.7 & --- & --- & 102.5 & 133.4 & --- \\
    Ap & --- & --- & 10.80 & 11.28 & --- \\
    DMDc & --- & --- & \textcolor{red}{\textbf{11.73}} & \textcolor{red}{\textbf{11.91}} & 0.00 / 0.00 \\
    \midrule
    \multirow{3}{*}{SINDy-AR} & 0.001 & 10.0 & \textcolor{blue}{\textbf{7.57}} & \textcolor{blue}{\textbf{7.68}} & 35.5 / 35.5 \\
     & 0.005 & 0.1 & 8.02 & 7.97 & 31.6 / 33.1 \\
     & 0.01 & 10.0 & 7.59 & 7.89 & 35.3 / 33.8 \\
    \bottomrule
  \end{tabular}
\end{table}

This same MAPE improvement can be observed by looking at the 5-day dynamic forecasting Tables \ref{tab:summary_5d_cae_p2} and \ref{tab:summary_5d_cae_p3}. In this situation, the improvement is even more marked as it reaches about 18\% for the training dataset. As previously observed, the test dataset relative percent improvement is less pronounced and reaches about 9\%, while it reached about 7.8\% for the 3-day case.

\begin{table}[htbp]
  \centering
  \caption{COAE-TIEGCM emulation summary, polynomial order 3 basis functions. MAPE (\%), 5-day propagation window. Training: 2002--2009. Test: 1996--2001.}
  \label{tab:summary_5d_cae_p3}
  \begin{tabular}{llcccc}
    \toprule
    Model & $\lambda$ & $\alpha_{\text{ridge}}$ & Train Avg (2002--2009) & Test Avg (1996--2001) & RMI vs.\ DMDc (\%) \\
    \midrule
    F10.7 & --- & --- & 102.5 & 133.4 & --- \\
    Ap & --- & --- & 10.80 & 11.28 & --- \\
    DMDc & --- & --- & \textcolor{red}{\textbf{11.73}} & \textcolor{red}{\textbf{11.91}} & 0.00 / 0.00 \\
    \midrule
    \multirow{3}{*}{SINDy-AR} & 0.001 & 10.0 & \textcolor{blue}{\textbf{6.18}} & \textcolor{blue}{\textbf{6.92}} & 47.3 / 41.9 \\
     & 0.005 & 0.1 & 6.38 & 6.86 & 45.6 / 42.4 \\
     & 0.01 & 10.0 & 6.60 & 7.18 & 43.7 / 39.7 \\
    \bottomrule
  \end{tabular}
\end{table}

Tables \ref{tab:summary_3d_tiegcm} and \ref{tab:summary_5d_tiegcm} summarize the emulation fidelity of the SINDy-AR framework in learning the TIEGCM dynamics for the 3 and 5 day dynamic propagation intervals in terms of MAPE and RMI of SINDy with respect to DMD (defined in Section \ref{sec:metrics}). Models are trained and tested on different solar cycles, thereby demonstrating a good generalization ability to out of sample dynamics.

Changing the threshold can also be interpreted as a way to move around basis functions. In fact, when the threshold is increased, some of the regression coefficients selected at the previous threshold typically become irrelevant, and the corresponding variables are ruled out.

The Tables show that the best SINDy-AR models outperform the baseline (DMDc) models every year. The overall improvement, defined in Section \ref{sec:metrics} as the absolute relative percent variation between the best SINDy-AR model and DMDc, is $\sim20\%$ for the 3 day emulation and $\sim 24\%$ for the 5 day propagation, showing consistency between the two experiments. 

Interestingly enough, the STLSQ threshold of 0.1 corresponds to a SINDy-AR model that ruled out enough features such that the resulting model is much closer to the DMDc model than the SINDy-AR model corresponding to STLSQ threshold of 0.001.

\subsection{Geomagnetic Storms}
\label{sec:geomagnetic_storms}
To study the behavior of the SINDy-AR framework under geomagnetic storms, during which the nonlinear regime of the thermosphere becomes most explicit, an experiment was developed so that the dynamic propagation during storms includes all periods when $A_p \geq 48$ ($k_p \geq 5$). Storm propagation starts 24 hours before the inception of the storm, and stops 30 hours after the end of the storm.

\begin{table}[htbp]
  \centering
  \caption{TIEGCM emulation. Summary MAPE (\%), storm-period propagation window. Training: 2002--2009. Test: 1996--2001.}
  \label{tab:summary_storm_tiegcm}
  \begin{tabular}{llcccc}
    \toprule
    Model & $\lambda$ & $\alpha_{\text{ridge}}$ & Train Avg (2002--2009) & Test Avg (1996--2001) & RMI vs.\ DMDc (\%) \\
    \midrule
    F10.7 & --- & --- & 102.5 & 133.4 & --- \\
    Ap & --- & --- & 10.93 & 11.32 & --- \\
    DMDc & --- & --- & \textcolor{red}{\textbf{23.05}} & \textcolor{red}{\textbf{17.71}} & 0.00 / 0.00 \\
    \midrule
    \multirow{5}{*}{SINDy-AR} & 0.001 & 10.0 & \textcolor{blue}{\textbf{8.09}} & \textcolor{blue}{\textbf{9.17}} & 64.9 / 48.2 \\
     & 0.005 & 0.1 & 8.41 & 9.28 & 63.5 / 47.6 \\
     & 0.01 & 10.0 & 9.30 & 10.48 & 59.7 / 40.8 \\
     & 0.05 & 0.1 & 19.94 & 21.81 & 13.5 / $-$23.2 \\
     & 0.1 & 10.0 & 23.88 & 20.93 & $-$3.6 / $-$18.2 \\
    \bottomrule
  \end{tabular}
\end{table}

\begin{table}[htbp]
  \centering
  \caption{COAE-TIEGCM emulation summary, polynomial order 2 basis functions. MAPE (\%), storm-period propagation window. Training: 2002--2009. Test: 1996--2001.}
  \label{tab:summary_storm_cae_p2}
  \begin{tabular}{llcccc}
    \toprule
    Model & $\lambda$ & $\alpha_{\text{ridge}}$ & Train Avg (2002--2009) & Test Avg (1996--2001) & RMI vs.\ DMDc (\%) \\
    \midrule
    F10.7 & --- & --- & 103.2 & 136.2 & --- \\
    Ap & --- & --- & 26.50 & 27.48 & --- \\
    DMDc & --- & --- & \textcolor{red}{\textbf{15.63}} & \textcolor{red}{\textbf{15.95}} & 0.00 / 0.00 \\
    \midrule
    \multirow{3}{*}{SINDy-AR} & 0.001 & 10.0 & \textcolor{blue}{\textbf{12.58}} & \textcolor{blue}{\textbf{12.18}} & 19.5 / 23.6 \\
     & 0.005 & 10.0 & 13.85 & 13.09 & 11.4 / 17.9 \\
     & 0.01 & 10.0 & 12.55 & 11.90 & 19.7 / 25.4 \\
    \bottomrule
  \end{tabular}
\end{table}

\begin{table}[htbp]
  \centering
  \caption{COAE-TIEGCM emulation summary, polynomial order 3 basis functions. MAPE (\%), storm-period propagation window. Training: 2002--2009. Test: 1996--2001.}
  \label{tab:summary_storm_cae_p3}
  \begin{tabular}{llcccc}
    \toprule
    Model & $\lambda$ & $\alpha_{\text{ridge}}$ & Train Avg (2002--2009) & Test Avg (1996--2001) & RMI vs.\ DMDc (\%) \\
    \midrule
    F10.7 & --- & --- & 103.2 & 136.2 & --- \\
    Ap & --- & --- & 26.50 & 27.48 & --- \\
    DMDc & --- & --- & \textcolor{red}{\textbf{15.63}} & \textcolor{red}{\textbf{15.95}} & 0.00 / 0.00 \\
    \midrule
    \multirow{3}{*}{SINDy-AR} & 0.001 & 10.0 & \textcolor{blue}{\textbf{9.87}} & \textcolor{blue}{\textbf{11.25}} & 36.9 / 29.5 \\
     & 0.005 & 10.0 & 10.15 & 11.19 & 35.1 / 29.8 \\
     & 0.01 & 10.0 & 10.36 & 11.17 & 33.7 / 30.0 \\
    \bottomrule
  \end{tabular}
\end{table}

The experiment was conducted using the linear (PCA) reduction method and the (nonlinear) COAE reduction method. We study the PCA case first.
Table \ref{tab:summary_storm_tiegcm} shows that the DMDc baseline results in a much larger MAPE value in the nonlinear regime than in the aggregated case. This is not surprising since the DMDc model is not suitable to describe nonlinearities, and, in fact, the MAPE of the best SINDy-AR model is between 2 and 3 times lower than DMDc. 

We observe that the absolute relative percent MAPE improvement between the DMDc baseline and the best SINDy-AR model during geomagnetic storms is between $\sim 48\%$ and $\sim 65\%$, which is almost three times the relative percent MAPE improvement calculated during all geomagnetic conditions which, as reported in Section \ref{sec:global_results}, is between $\sim 20\%$ and $\sim 24\%$. 

During storms, the SINDy-AR framework learns a better emulator from the PCA than from the COAE framework, similarly to what happens during quiet periods. A potential explanation is the ``incompatibility" of the SINDy-AR nonlinear regression framework with the neural network framework used by the COAE. We still observe from Tables \ref{tab:summary_storm_cae_p2} and \ref{tab:summary_storm_cae_p3} that increasing the order of nonlinearity of the basis functions improves the situation, but PCA+SINDy-AR still beats COAE+SINDy-AR also during storms.

\begin{figure}[h!]
\centering

\begin{subfigure}[t]{1\textwidth}
    \centering
    \includegraphics[width=\textwidth]{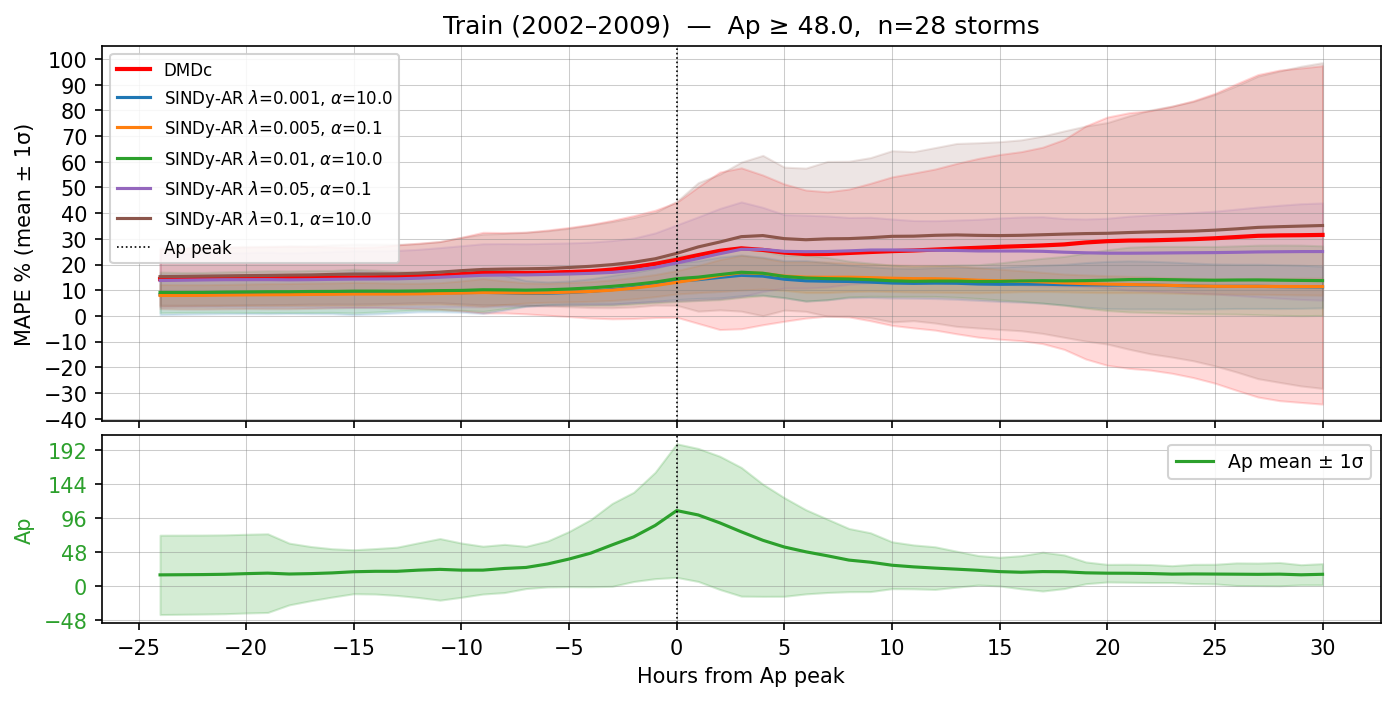}
    \caption{PCA-TIEGCM superposed epoch analysis. Training dataset.}
    \label{fig:tiegcm_superposed_pca_train}
\end{subfigure}
\hfill
\begin{subfigure}[t]{1\textwidth}
    \centering
    \includegraphics[width=\textwidth]{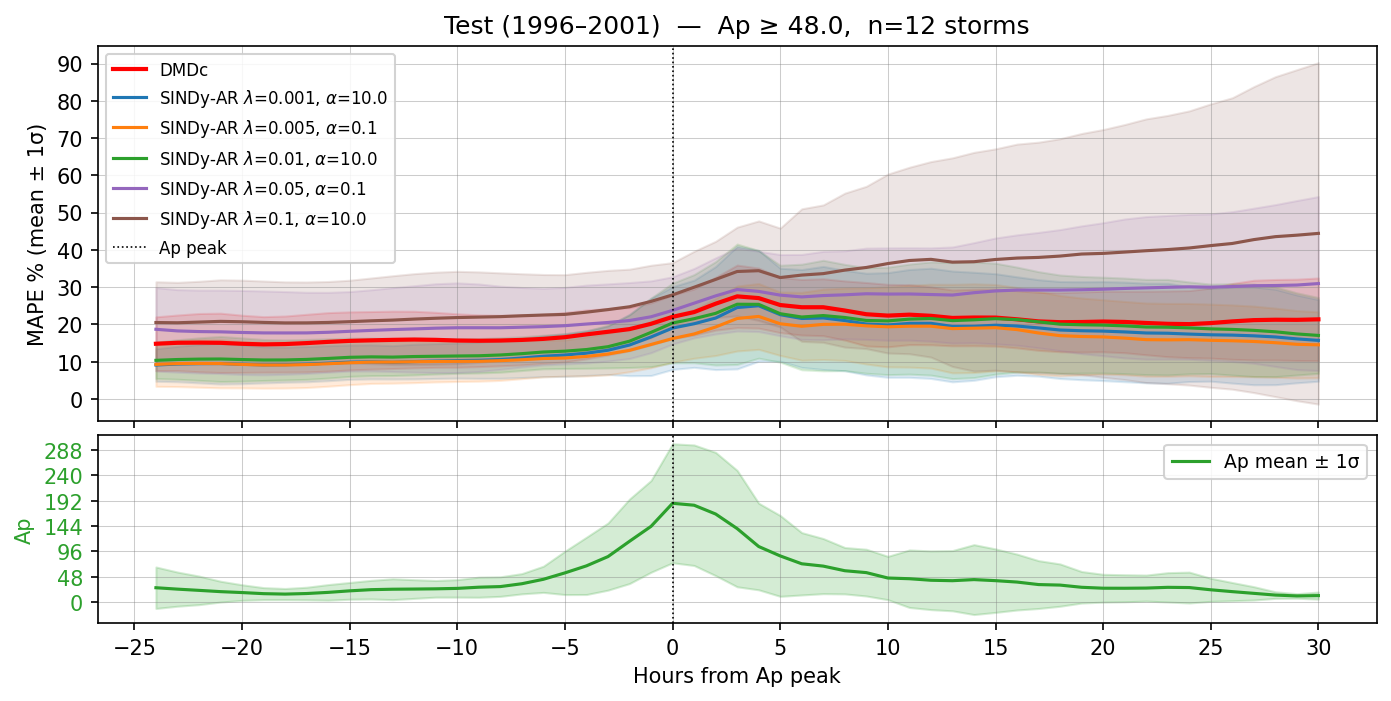}
    \caption{PCA-TIEGCM superposed epoch analysis. Test dataset.}
    \label{fig:tiegcm_superposed_pca_test}
\end{subfigure}

\caption{Superposed epoch analysis during geomagnetic storms ($A_p \geq 5$). Polynomial basis functions of order 2.}
\label{fig:all_full_superposed}

\end{figure}

\begin{figure}[h!]
\centering

\begin{subfigure}[t]{1\textwidth}
    \centering
    \includegraphics[width=\textwidth]{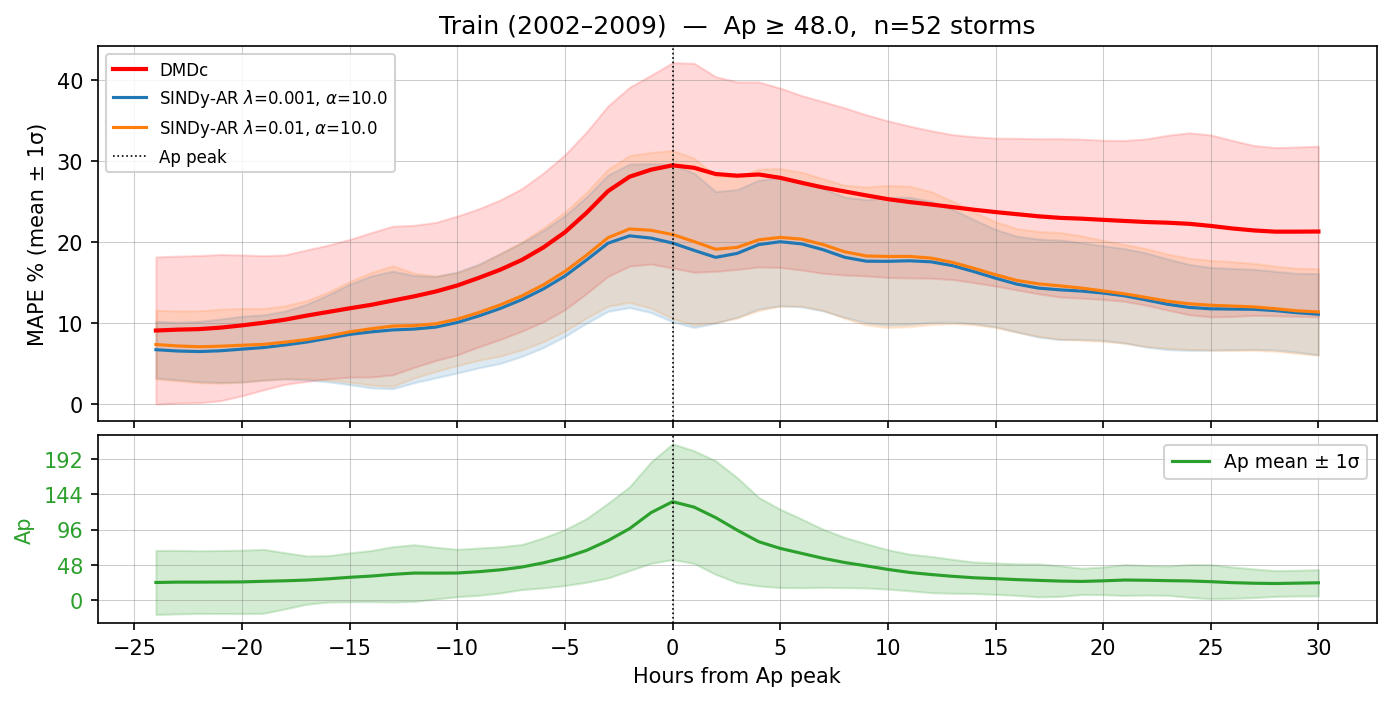}
    \caption{COAE-TIEGCM superposed epoch analysis. Training dataset.}
    \label{fig:tiegcm_superposed_coae_train}
\end{subfigure}
\hfill
\begin{subfigure}[t]{1\textwidth}
    \centering
    \includegraphics[width=\textwidth]{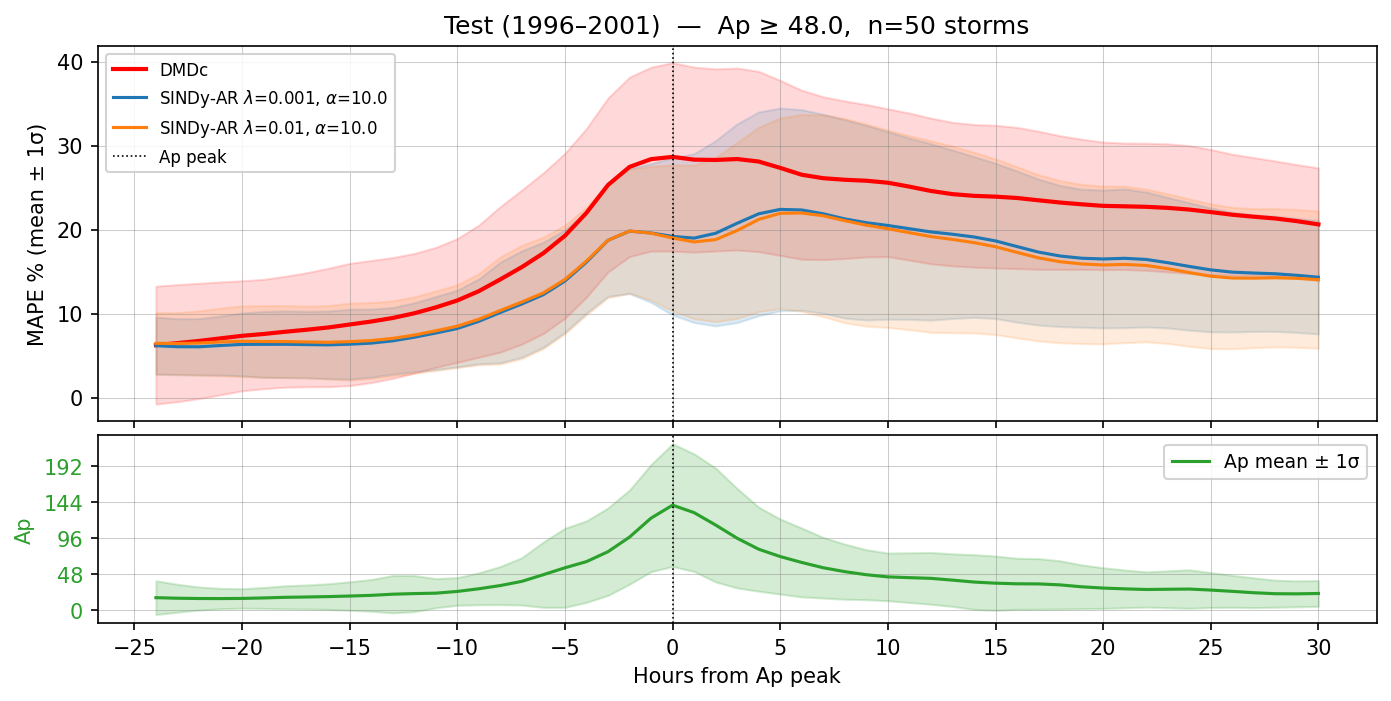}
    \caption{COAE-TIEGCM superposed epoch analysis. Test dataset.}
    \label{fig:tiegcm_superposed_coae_test}
\end{subfigure}

\caption{Superposed epoch analysis during geomagnetic storms ($A_p \geq 5$). Polynomial basis functions of order 3.}
\label{fig:all_full_superposed_coae}

\end{figure}

Further information can be found from the superposed epoch analysis in Figures \ref{fig:all_full_superposed} and \ref{fig:all_full_superposed_coae}, showing the PCA and COAE versions respectively. In these plots, the MAPE of every storm propagation is collected for every class of hyperparameters, and the average and standard deviations of MAPE at every hour are calculated. The lower subplots in Figures \ref{fig:tiegcm_superposed_pca_train}, \ref{fig:tiegcm_superposed_pca_test}, \ref{fig:tiegcm_superposed_coae_train} and \ref{fig:tiegcm_superposed_coae_test} show the superposed analysis of $A_p$ through its mean and standard deviation.

The COAE version of the superposed epoch analysis, shown in Figure \ref{fig:all_full_superposed_coae}, shows something compatible with what previously said. However, the difference between the average value of the MAPE of the SIND-AR models with that of the DMD model is more neatly pronounced than in the PCA case. As DMD is an inherently linear model, this supports the thesis a highly nonlinear latent space is ``hard" to emulate with linear techniques. We should also take into account that this COAE version mainly focuses on minimizing the reconstruction error in an orthogonal framework \cite{rope_dim_red}. Hence, an open question is how much this nonlinear representation inability would improve when explicitly considering the dynamics of the underlying thermosphere. 

Table \ref{tab:summary_storm_tiegcm} highlights the need to build a nonlinear framework from DMDc to better emulate the thermosphere, which was achieved with the SINDy-AR. The DMDc MAPE stabilizes around an average value of 20\% after reaching a maximum close to 30\% around $A_p = 48$, and the best SINDy-AR model tends to go back to 8.5\% after reaching a peak of $\sim 20\%$ around the storm peak. We observe that the MAPE peak follows the $A_p$ peak by 2 to 4 hours. In general, the SINDy-AR models, although consistently improving over DMDc during all the phases of the storm, experience some delays in reaching the stationary state again during the recovery phase. This observation raises questions about potential new variables and techniques to use in future applications. This discussion is addressed in Section \ref{sec:conclusion}.
Figure \ref{fig:all_full_superposed} shows that the SINDy-AR models span different MAPE values during storm recovery. We found that DMDc is systematically worse at any stage of the storm with respect of most of the SINDy-AR models. However, although autoregressive steps and memory variables are close to optimal from experiments, one of the Achilles heels of the SINDy-AR models shows up during the recovery phase. We suggest some improvements to apply in Section \ref{sec:conclusion}.

\subsection{Implementing the Framework in Practice: Propagation Times and Forecast Fidelity}
\label{sec:prop_times}

\begin{figure}[h!]
\centering
\begin{subfigure}[t]{1\textwidth}
    \centering
    \includegraphics[width=\textwidth]{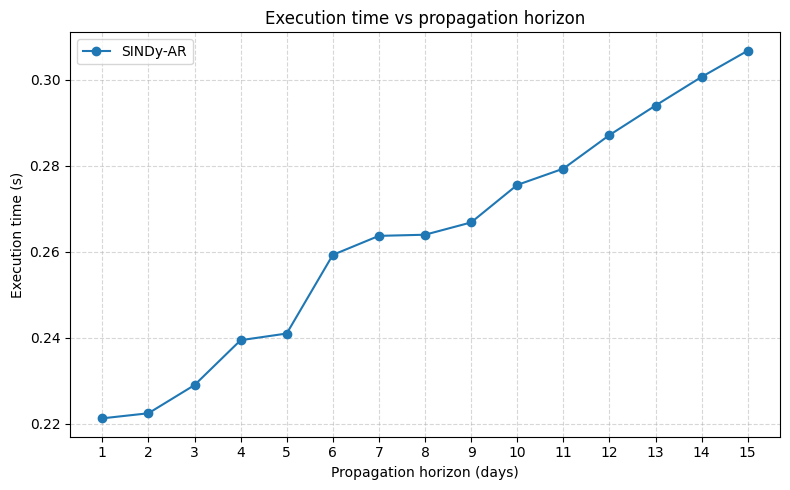}
    \caption{Propagation.}
    \label{fig:prop_time}
\end{subfigure}
\hfill
\begin{subfigure}[t]{1\textwidth}
    \centering
    \includegraphics[width=\textwidth]{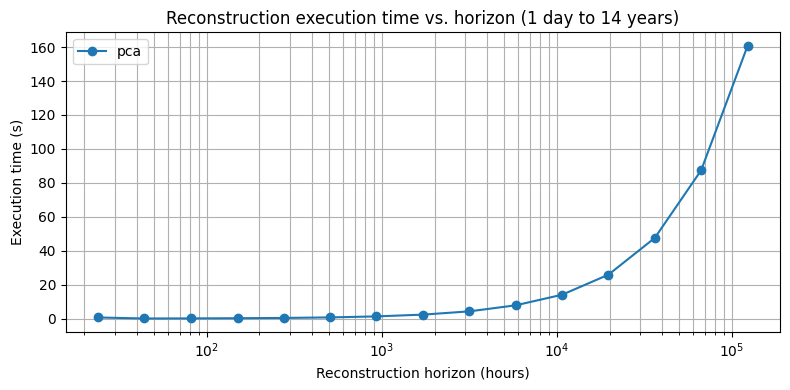}
    \caption{Reconstruction.}
    \label{fig:recon_time}
\end{subfigure}
\hfill
\begin{subfigure}[t]{1\textwidth}
    \centering
    \includegraphics[width=\textwidth]{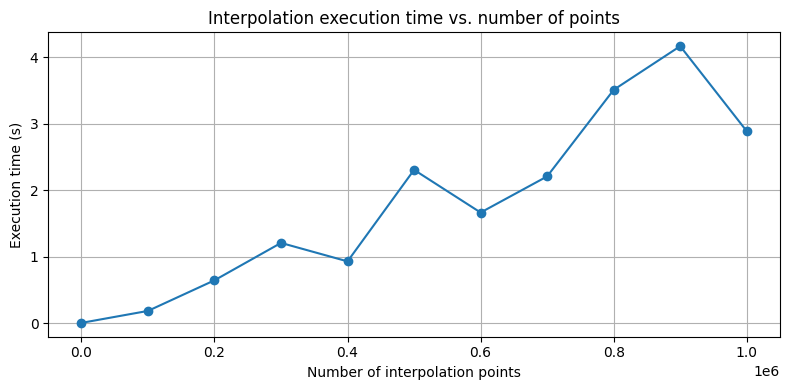}
    \caption{Interpolation.}
    \label{fig:interp_time}
\end{subfigure}

\caption{Analysis of execution times for propagation, reconstruction and interpolation.}
\label{fig:execution_times}
\end{figure}

A system that can be set up on board a satellite for situational awareness purposes will need to be as fast as possible for real time calculations and have the smallest possible power consumption. Hand in hand with this requirement, the system needs to be accurate enough to calculate trajectory corrections. The execution times of the framework can be found in Figure \ref{fig:execution_times}, showing the time required to propagate the system in the global 10 dimensional latent space. Results indicate that the trend is almost linear, and a 4 day propagation of the global state of the thermosphere for a single model in the latent space requires only 0.24 seconds. 
Some operators may require only a propagation of a few hours, which means that this execution time is even smaller in those cases. The computational cost of SINDy-AR versus TIEGCM is found by calculating the ratio of the execution rates of the two models across the 14 year period.
Regarding TIEGCM, two weeks of calculation for one year of simulation implies a rate of

\begin{align}
    v_{\text{TIEGCM}} = \frac{365 \cdot 24 \cdot 60 \cdot 60 sec}{14 \cdot 24 \cdot 60 \cdot 60 sec}\sim 26.07
\end{align}

which is obtained by simulating on HPC. The SINDy-AR execution rate is obtained through

\begin{align}
    v_{\text{SINDy-AR}} = \frac{14 \cdot 365 \cdot 24 \cdot 60 \cdot 60 sec}{31 sec}\sim 14,242,064,
\end{align}

which is the rate of simulation in the latent space. This is about $550,000$ times faster than the TIEGCM runtime on HPC. However, reconstructing the global density grid of 116,640 points across the 14-year horizon takes another 157 seconds total, as shown in Figure \ref{fig:recon_time}, for a total runtime that becomes $90,000$ times faster than the original TIEGCM runtime (still compared to the TIEGCM HPC execution times).

After reconstruction to measurement space ($\frac{kg}{m^3}$), one useful application is to interpolate at the spatial locations where the density is required. Since the reduced order method used here is PCA (see Equation \ref{eq:pca}), a method that allows for the pre-interpolation of the reconstruction matrices, interpolation becomes much faster than standard interpolation, resulting in very low reconstruction times. 

The system currently allows for the interpolation of multiple points at the same time, which makes the framework particularly well suited for situational awareness involving mega-constellations of satellites. Figure \ref{fig:interp_time} shows that calculating the density at the locations of 1,000,000 satellites requires 3 seconds. The densities at the locations of 400,000 satellites require only one second of calculations, and the densities at the locations of 10,000 satellites require virtually no time at all.

\begin{figure}[h]
\centering
\includegraphics[width=1\textwidth]{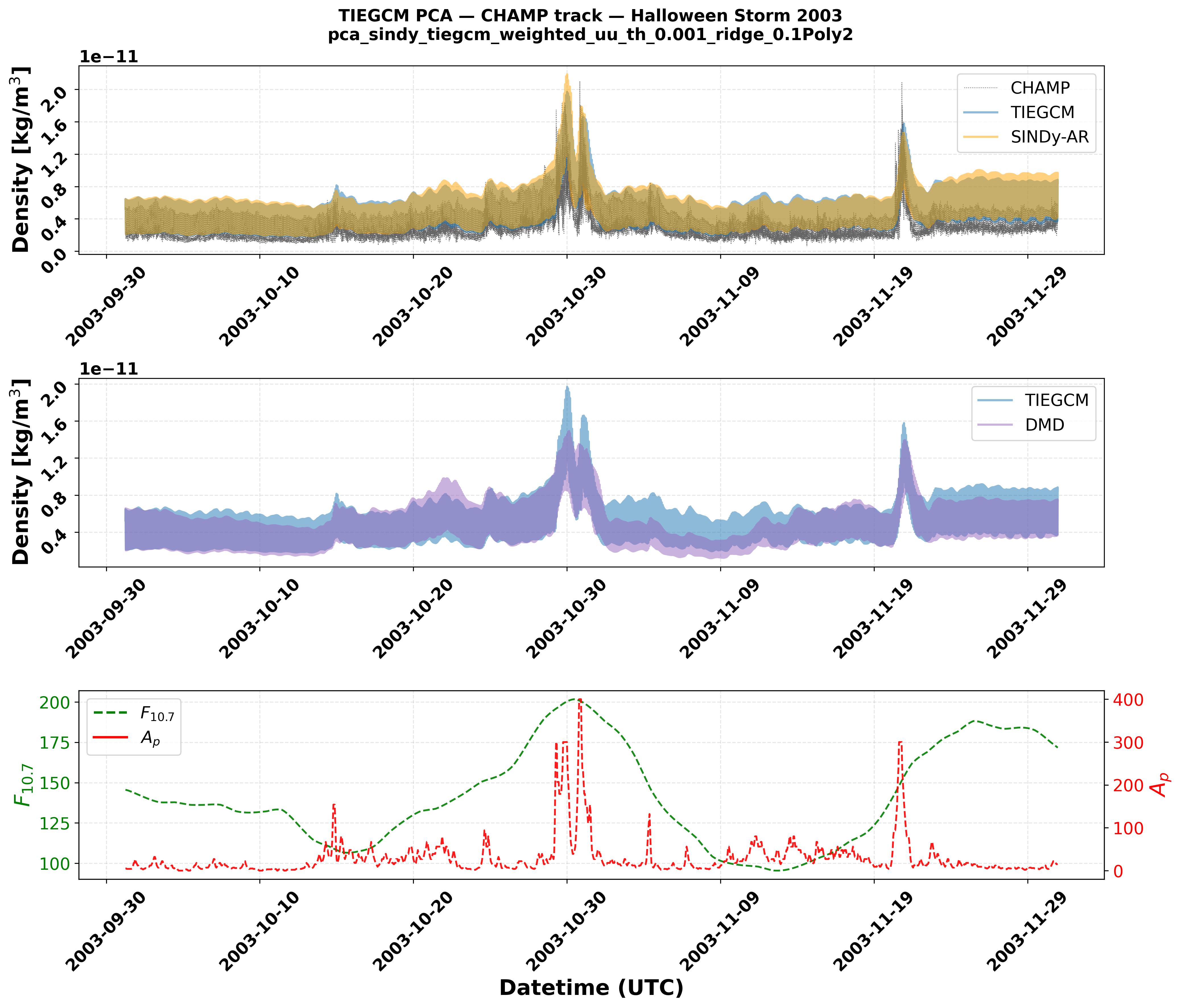}
\caption{Interpolation along the CHAMP trajectory during the Halloween Storm. The first subplot includes the accelerometer derived density of the CHAMP satellite. The improvement in accuracy between DMD and SINDy-AR in emulating TIEGCM is visually noticeable.}
\label{fig:halloween_storm_dmd_validation}
\end{figure}

Figure \ref{fig:halloween_storm_dmd_validation} validates the SINDy-AR density of the best model (STLSQ = 0.001 and ridge = 10), represented along the TIEGCM and the CHAMP accelerometer density in the first subplot, against DMDc and TIEGCM in the second subplot, thus showing their differences. The analysis is done using data from October $1^{\text{st}}$ 2003 to November $30^{\text{th}}$ 2003 to include the Halloween storm, which is arguably one of the major storms that have occurred up to now. The third subplot shows solar drivers and how they evolve during the storm. The emulation fidelity of SINDy-AR versus DMDc can be grasped visually at first. However, error metrics were collected and indicate a 17.73\% MAPE for DMDc and a 5.21\% MAPE for SINDy-AR with respect to TIEGCM over those two months along the CHAMP orbit. The accelerometer density is added only for visual comparison purposes; further work is required to evaluate the performance of the SINDy-AR model relative to the CHAMP density. In addition, the Pearson correlation coefficient between the SINDy-AR density and the CHAMP density during the 2-month Halloween storm propagation is 89.32\%. The same correlation coefficient between the original TIEGCM and CHAMP during the same period is 89.50\%, thus showing that the physics-based model generates dynamics that are highly correlated to the observed density, and that this dynamical behavior can be very accurately captured by the learned SINDy-AR models.

\section{Conclusion and future work}
\label{sec:conclusion}

This work represents the first application of the SINDy-AR framework using STLSQ and stability analysis in a high dimensional system such as the thermosphere, thus improving and complementing the results obtained in \cite{sindy0}. This is a result with potential for the community for scientific and operational applications.

The Article also describes how to efficiently deploy a selection of stable SINDy-AR thermospheric emulators to operations from which model uncertainty can be built through the calculation of the mean and the standard deviation.

The work addresses how to deal with the largest source of uncertainty in operations in LEO by building a fast, uncertainty-aware SINDy-AR framework to emulate thermospheric density for space situational awareness and scientific applications. Current physics-based models require large computational times. SINDy-AR runs orders of magnitude faster than the underlying physics-based model (see Section \ref{sec:prop_times}). SINDy-AR is capable to achieve a forecast fidelity similar to that of the underlying physics-based model, with an accuracy that can be quantified using the results shown in Section \ref{sec:global_results}. SINDy-AR is fast, shows higher fidelity to emulate TIEGCM than what can be done using DMDc, and can also deliver uncertainty quantification quickly as a natural byproduct of its outputs, as opposed to physics-based and empirical models.

This study also complements the results obtained in Paper \cite{rope_dim_red_2}. Remarkably, the PCA based SINDy-AR models with STLSQ thresholding and stability requirements have performances that are comparable to or only slightly less accurate than the best ensemble of models reported in Table 3 of Paper \cite{rope_dim_red_2}. The ensemble in \cite{rope_dim_red_2} is particularly more accurate during the quiet years 2006, 2007, 2008, and 2009, which are also part of the training dataset. During the test years, 1996 (validation in \cite{rope_dim_red_2} and test here) through 2001, the SINDy-AR models are comparable to the neural network ensemble, being slightly more accurate in years like 1999, 2000 (validation in \cite{rope_dim_red_2}), and 2001, and slightly less accurate in years such as 1996, 1997, and 1998.

We showed that the SINDy-AR framework can deliver actionable and uncertainty aware results for operations, taking virtually no time to propagate the global state of the density for a few hours or even a couple of days and interpolate to get the values of interest along the trajectories of thousands of satellites. Figure \ref{fig:halloween_storm_dmd_validation} presents this result and serves as a test case for operations. It shows the propagation of the best SINDy-AR model of the framework across October and November 2003, the period including the Halloween Storm, along the trajectory of the CHAMP satellite. Propagation and evaluation during this time period require execution times of only a few seconds. The same interval is used to benchmark the learned SINDy-AR, along the CHAMP trajectory to show how it improves with respect to nonlinear-DMDc. The validation demonstrates that there is an important difference between the SINDy-AR and CHAMP densities as well as between the TIEGCM and CHAMP densities. However, further work is required to understand how to remove that difference at different locations and times.

As a side observation of the experiments, a few of the models that were ruled out because they were unstable showed higher accuracy than the best stable models during geomagnetic storms. The phase space of such models was not studied. However, future work can go in the direction of finding subspaces that can be considered ATRs in terms of the analysis of \cite{atr_stability}. Once conditionally stable models are better understood, they can be used whenever they demonstrate superior accuracy. The resulting output would be a model that is ‘‘projected'' onto a hybrid ATR whenever an otherwise more accurate model threatens to leave it. The decision of which model to pick, at which point of the dynamics, and with what significance can be made using a fuzzy logic controller, for instance.

Although most of the nonlinear models are systematically better than DMDc during storms (see Figure \ref{fig:all_full_superposed}), Section \ref{sec:geomagnetic_storms} identifies one of the Achilles heels of the SINDy-AR framework during the recovery phase. The recovery phase accuracy can be improved for instance by taking into account a storm phase feature. It may be $A_p$ time rate of change or time since last peak. This can be used as a characterization of how long the density stays up after $A_p$ drops. A 48-hour rolling standard deviation of $A_p$ is already being used as an external drivers to keep into account this factor. However, such alternatives to the standard deviation of $A_p$ can be tested in a future work. Moreover, the 5-hour AR step has been tuned to span the tidal content, not the transient decay that follows a storm. In this way, a single stable model can learn phase-dependent behavior instead of a single set of coefficients covering quiet and recovery dynamics equally. Since the training set (2002-2009) includes plenty of storms, an Ap-triggered regime switch to a second, more accurate but only conditionally stable model during the storm and recovery window, falling back to the guaranteed-stable model otherwise, is a direct extension of what is already built. 

The SINDy-AR framework was built to study, learn, and leverage physics-based models of high dimensional systems in applications such as space weather and operations, in particular with ROPE, and it was inspired by \cite{sindy_pnas}.

\section*{Open Research Section}
\paragraph*{Data Availability Statement}
\label{sec:data_availability}

The code, together with a working set of models, the analyses, and the figures presented in this manuscript is publicly available on Zenodo under a CC~BY~4.0 license at: \url{https://doi.org/10.5281/zenodo.20597781}. This includes part of the models and the model coefficients derived in the PCA latent space. The .mat files also include the latent space dataset and space weather drivers that have been used.

The model training data are outputs of the NCAR Thermosphere-Ionosphere-Electrodynamics General Circulation Model (TIEGCM), provided by Dr. Eric Sutton. The model coefficients (.mat files) encode a PCA+SINDy-AR representation trained on TIEGCM simulations driven by observed solar and geomagnetic indices (1996–2009).

The CHAMP data can be found at \url{http://thermosphere.tudelft.nl/}.

\paragraph*{Conflict of Interest Statement}
The authors have no conflicts of interest to disclose.

\acknowledgments
This research is based upon work support in part by the Office of the Director of National Intelligence (ODNI), Intelligence Advance Research Projects Activity (IARPA), via 2023–23060200005. The views and conclusions contained herein are those of the authors and should not be interpreted as necessarily representing official policies, either expressed or implied, of ODNI, IARPA, or the U.S. Government. The U.S. Government is authorized to reproduce and distribute reprints for governmental purposes notwithstanding any copyright annotation therein. This work was also supported in part by a project funded through the University of Colorado Boulder (OSC via CU CIRES). The authors also thank the members of the Astrodynamics, Space Science \& Space Technology (ASSIST) lab at WVU for their support.
PMM also gratefully acknowledges support under NSF CAREER Award \#2140204 and NSF ANSWERS Award \#2149747 (sub award to West Virginia University from Rutgers University). 

\bibliography{agusample}   

\appendix
\section{Detailed Tables}
Here we provide tables with yearly details across the dataset.
\renewcommand{\thetable}{A\arabic{table}}
\setcounter{table}{0}

\begin{sidewaystable}
  \centering
  \caption{TIEGCM emulation. Mean MAPE (\%) by year for a 3-day propagation window. Training is performed from 2002 to 2009, test is from 1996 to 2001. Propagation is reset every 3 days. Each propagation interval starts by using the actual density, transforming it to the initial latent state, then propagating it in the latent space. Dynamic forecasting uses the previously forecast output as the initial condition to the next propagation.}
  \label{tab:mape_year_3d_pca_tiegcm}
  \resizebox{\textwidth}{!}{%
  \begin{tabular}{lllrrrrrrrrr|rrrrrrr}
    \toprule
    \multicolumn{3}{c}{} & \multicolumn{9}{c}{Train} & \multicolumn{7}{c}{Test} \\
    \cmidrule(lr){4-12} \cmidrule(lr){13-19}
    Model & $\lambda$ & $\alpha_{\text{ridge}}$ & 2002 & 2003 & 2004 & 2005 & 2006 & 2007 & 2008 & 2009 & Avg & 1996 & 1997 & 1998 & 1999 & 2000 & 2001 & Avg \\
    \midrule
    F10.7 & --- & --- & 195.0 & 140.0 & 114.5 & 97.7 & 79.8 & 69.1 & 62.9 & 61.2 & 102.5 & 71.8 & 82.2 & 122.5 & 155.0 & 186.8 & 182.2 & 133.4 \\
    Ap & --- & --- & 12.79 & 21.20 & 13.04 & 13.07 & 8.30 & 7.31 & 6.78 & 3.85 & 10.79 & 8.56 & 8.13 & 11.69 & 12.20 & 14.70 & 12.51 & 11.30 \\
    \midrule
    DMDc & --- & --- & 8.30 & 8.71 & 7.71 & 7.98 & 6.90 & 6.91 & 7.10 & 6.32 & \textcolor{red}{\textbf{7.49}} & 6.51 & 6.74 & 8.11 & 8.02 & 9.25 & 9.02 & \textcolor{red}{\textbf{7.94}} \\
    \midrule
    \multirow{5}{*}{SINDy-AR} & 0.001 & 10.0 & 6.73 & 7.32 & 5.99 & 6.25 & 5.56 & 5.42 & 5.90 & 5.80 & \textcolor{blue}{\textbf{6.12}} & 5.36 & 5.45 & 6.33 & 6.05 & 6.97 & 6.91 & \textcolor{blue}{\textbf{6.18}} \\
     & 0.005 & 0.1 & 7.29 & 7.48 & 6.27 & 6.66 & 6.02 & 5.47 & 5.76 & 5.43 & 6.30 & 6.14 & 5.76 & 6.41 & 6.34 & 7.47 & 7.40 & 6.59 \\
     & 0.01 & 10.0 & 6.78 & 7.45 & 6.11 & 6.27 & 5.58 & 5.45 & 5.90 & 5.77 & 6.16 & 5.38 & 5.50 & 6.45 & 6.12 & 6.98 & 6.96 & 6.23 \\
     & 0.05 & 0.1 & 8.41 & 9.19 & 8.28 & 8.93 & 8.34 & 7.76 & 7.79 & 6.64 & 8.17 & 8.01 & 7.73 & 8.23 & 7.80 & 8.98 & 8.71 & 8.24 \\
     & 0.1 & 10.0 & 8.48 & 9.52 & 8.07 & 8.32 & 7.95 & 7.81 & 8.11 & 7.38 & 8.21 & 7.40 & 7.22 & 8.13 & 7.91 & 9.59 & 9.12 & 8.23 \\
    \bottomrule
  \end{tabular}
  }
\end{sidewaystable}

\begin{sidewaystable}
  \centering
  \caption{TIEGCM emulation. Mean MAPE (\%) by year for a 5-day propagation window. Training is performed from 2002 to 2009, test is from 1996 to 2001. Propagation is reset every 5 days. Each propagation interval starts by using the actual density, transforming it to the initial latent state, then propagating it in the latent space. Dynamic forecasting uses the previously forecast output as the initial condition to the next propagation.}
  \label{tab:mape_year_5d_pca_tiegcm}
  \resizebox{\textwidth}{!}{%
  \begin{tabular}{lllrrrrrrrrr|rrrrrrr}
    \toprule
    \multicolumn{3}{c}{} & \multicolumn{9}{c}{Train} & \multicolumn{7}{c}{Test} \\
    \cmidrule(lr){4-12} \cmidrule(lr){13-19}
    Model & $\lambda$ & $\alpha_{\text{ridge}}$ & 2002 & 2003 & 2004 & 2005 & 2006 & 2007 & 2008 & 2009 & Avg & 1996 & 1997 & 1998 & 1999 & 2000 & 2001 & Avg \\
    \midrule
    F10.7 & --- & --- & 195.0 & 140.0 & 114.5 & 97.7 & 79.7 & 69.1 & 62.9 & 61.2 & 102.5 & 71.8 & 82.2 & 122.5 & 155.0 & 186.8 & 182.2 & 133.4 \\
    Ap & --- & --- & 12.76 & 21.21 & 13.01 & 13.06 & 8.28 & 7.30 & 6.79 & 3.86 & 10.78 & 8.56 & 8.14 & 11.66 & 12.19 & 14.69 & 12.51 & 11.29 \\
    \midrule
    DMDc & --- & --- & 9.41 & 9.96 & 8.84 & 9.27 & 7.84 & 7.81 & 8.06 & 6.87 & \textcolor{red}{\textbf{8.51}} & 7.06 & 7.52 & 9.25 & 9.13 & 10.40 & 10.54 & \textcolor{red}{\textbf{8.99}} \\
    \midrule
    \multirow{5}{*}{SINDy-AR} & 0.001 & 10.0 & 7.29 & 7.95 & 6.67 & 6.71 & 5.96 & 5.65 & 6.29 & 6.01 & \textcolor{blue}{\textbf{6.57}} & 5.65 & 5.88 & 6.90 & 6.52 & 7.45 & 7.54 & \textcolor{blue}{\textbf{6.66}} \\
     & 0.005 & 0.1 & 7.88 & 7.93 & 6.90 & 7.04 & 6.62 & 5.91 & 6.36 & 5.72 & 6.79 & 6.72 & 6.25 & 6.78 & 6.66 & 7.87 & 8.10 & 7.06 \\
     & 0.01 & 10.0 & 7.36 & 8.07 & 6.85 & 6.79 & 6.08 & 5.81 & 6.42 & 6.08 & 6.68 & 5.75 & 6.02 & 7.05 & 6.56 & 7.57 & 7.72 & 6.78 \\
     & 0.05 & 0.1 & 9.66 & 9.65 & 9.01 & 9.89 & 9.72 & 8.66 & 9.06 & 7.23 & 9.11 & 9.24 & 8.40 & 8.68 & 8.29 & 10.00 & 9.67 & 9.05 \\
     & 0.1 & 10.0 & 9.52 & 10.26 & 9.31 & 9.29 & 9.07 & 8.76 & 9.35 & 8.19 & 9.22 & 8.26 & 8.00 & 8.73 & 8.80 & 11.43 & 10.89 & 9.35 \\
    \bottomrule
  \end{tabular}
  }
\end{sidewaystable}

\begin{sidewaystable}
  \centering
  \caption{TIEGCM emulation. Mean MAPE (\%) by year for the storm-period propagation window.}
  \label{tab:storms}
  \resizebox{\textwidth}{!}{%
  \begin{tabular}{lllrrrrrrrrr|rrrrrrr}
    \toprule
    \multicolumn{3}{c}{} & \multicolumn{9}{c}{Train} & \multicolumn{7}{c}{Test} \\
    \cmidrule(lr){4-12} \cmidrule(lr){13-19}
    Model & $\lambda$ & $\alpha_{\text{ridge}}$ & 2002 & 2003 & 2004 & 2005 & 2006 & 2007 & 2008 & 2009 & Avg & 1996 & 1997 & 1998 & 1999 & 2000 & 2001 & Avg \\
    \midrule
    F10.7 & --- & --- & 195.0 & 140.0 & 114.5 & 97.7 & 79.7 & 69.2 & 63.0 & 61.2 & 102.5 & 71.8 & 81.9 & 122.5 & 155.0 & 186.8 & 182.2 & 133.4 \\
    Ap & --- & --- & 12.78 & 21.27 & 13.08 & 13.12 & 8.46 & 7.53 & 7.03 & 4.21 & 10.93 & 8.57 & 8.26 & 11.68 & 12.21 & 14.69 & 12.54 & 11.32 \\
    \midrule
    DMDc & --- & --- & 28.49 & 65.06 & 28.24 & 15.50 & 12.19 & 10.58 & 13.58 & 10.76 & \textcolor{red}{\textbf{23.05}} & 11.09 & 13.74 & 17.73 & 20.16 & 22.01 & 21.51 & \textcolor{red}{\textbf{17.71}} \\
    \midrule
    \multirow{5}{*}{SINDy-AR} & 0.001 & 10.0 & 10.52 & 10.50 & 8.53 & 8.45 & 6.72 & 6.05 & 6.74 & 7.19 & \textcolor{blue}{\textbf{8.09}} & 6.59 & 7.39 & 9.85 & 9.24 & 10.70 & 11.23 & \textcolor{blue}{\textbf{9.17}} \\
     & 0.005 & 0.1 & 11.54 & 11.00 & 8.54 & 8.13 & 7.43 & 6.60 & 7.08 & 6.97 & 8.41 & 8.92 & 7.37 & 8.82 & 9.58 & 11.05 & 9.96 & 9.28 \\
     & 0.01 & 10.0 & 10.75 & 13.23 & 9.56 & 9.59 & 8.13 & 6.92 & 8.04 & 8.14 & 9.30 & 8.50 & 9.10 & 11.61 & 10.44 & 11.08 & 12.16 & 10.48 \\
     & 0.05 & 0.1 & 25.59 & 31.04 & 25.41 & 20.60 & 14.74 & 12.06 & 17.13 & 12.96 & 19.94 & 18.04 & 20.51 & 19.14 & 23.03 & 27.97 & 22.21 & 21.81 \\
     & 0.1 & 10.0 & 29.34 & 34.55 & 36.74 & 27.85 & 14.54 & 12.85 & 19.14 & 16.05 & 23.88 & 14.85 & 20.23 & 21.10 & 18.89 & 26.75 & 23.78 & 20.93 \\
    \bottomrule
  \end{tabular}
  }
\end{sidewaystable}

\begin{sidewaystable}[htbp]
  \centering
  \caption{COAE-TIEGCM emulation, polynomial of order 2 basis functions. All weather conditions. Mean MAPE (\%) by year for the 3-day propagation window.}
  \label{tab:mape_year_3d_cae_poly2}
  \resizebox{\textwidth}{!}{%
  \begin{tabular}{lllrrrrrrrrr|rrrrrrr}
    \toprule
    \multicolumn{3}{c}{} & \multicolumn{9}{c}{Train} & \multicolumn{7}{c}{Test} \\
    \cmidrule(lr){4-12} \cmidrule(lr){13-19}
    Model & $\lambda$ & $\alpha_{\text{ridge}}$ & 2002 & 2003 & 2004 & 2005 & 2006 & 2007 & 2008 & 2009 & Avg & 1996 & 1997 & 1998 & 1999 & 2000 & 2001 & Avg \\
    \midrule
    F10.7 & --- & --- & 195.0 & 140.0 & 114.5 & 97.7 & 79.8 & 69.1 & 62.9 & 61.2 & 102.5 & 71.8 & 82.2 & 122.4 & 155.1 & 186.8 & 182.2 & 133.4 \\
    Ap & --- & --- & 12.78 & 21.25 & 13.07 & 13.09 & 8.30 & 7.33 & 6.78 & 3.86 & 10.81 & 8.55 & 8.14 & 11.67 & 12.18 & 14.69 & 12.53 & 11.29 \\
    \midrule
    DMDc & --- & --- & 9.94 & 11.56 & 9.58 & 9.83 & 9.45 & 9.35 & 9.83 & 8.84 & \textcolor{red}{\textbf{9.80}} & 9.16 & 8.51 & 10.44 & 10.21 & 11.30 & 10.66 & \textcolor{red}{\textbf{10.05}} \\
    \midrule
    \multirow{3}{*}{SINDy-AR} & 0.001 & 10.0 & 6.79 & 7.25 & 6.78 & 7.38 & 6.63 & 6.51 & 6.36 & 5.31 & \textcolor{blue}{\textbf{6.62}} & 6.26 & 6.08 & 6.83 & 6.62 & 7.49 & 7.56 & \textcolor{blue}{\textbf{6.81}} \\
     & 0.005 & 0.1 & 7.18 & 7.71 & 7.50 & 8.04 & 6.89 & 6.62 & 6.50 & 5.35 & 6.97 & 6.62 & 6.36 & 7.15 & 6.63 & 7.35 & 7.35 & 6.91 \\
     & 0.01 & 10.0 & 6.66 & 7.56 & 6.94 & 7.56 & 6.59 & 6.38 & 6.30 & 5.20 & 6.65 & 6.23 & 6.18 & 6.94 & 6.66 & 7.47 & 7.02 & 6.75 \\
    \bottomrule
  \end{tabular}
  }
\end{sidewaystable}

\begin{sidewaystable}[htbp]
  \centering
  \caption{COAE-TIEGCM emulation, polynomial of order 3 basis functions. All weather conditions. Mean MAPE (\%) by year for the 3-day propagation window.}
  \label{tab:mape_year_3d_cae_poly3}
  \resizebox{\textwidth}{!}{%
  \begin{tabular}{lllrrrrrrrrr|rrrrrrr}
    \toprule
    \multicolumn{3}{c}{} & \multicolumn{9}{c}{Train} & \multicolumn{7}{c}{Test} \\
    \cmidrule(lr){4-12} \cmidrule(lr){13-19}
    Model & $\lambda$ & $\alpha_{\text{ridge}}$ & 2002 & 2003 & 2004 & 2005 & 2006 & 2007 & 2008 & 2009 & Avg & 1996 & 1997 & 1998 & 1999 & 2000 & 2001 & Avg \\
    \midrule
    F10.7 & --- & --- & 195.0 & 140.0 & 114.5 & 97.7 & 79.8 & 69.1 & 62.9 & 61.2 & 102.5 & 71.8 & 82.2 & 122.4 & 155.1 & 186.8 & 182.2 & 133.4 \\
    Ap & --- & --- & 12.78 & 21.25 & 13.07 & 13.09 & 8.30 & 7.33 & 6.78 & 3.86 & 10.81 & 8.55 & 8.14 & 11.67 & 12.18 & 14.69 & 12.53 & 11.29 \\
    \midrule
    DMDc & --- & --- & 9.94 & 11.56 & 9.58 & 9.83 & 9.45 & 9.35 & 9.83 & 8.84 & \textcolor{red}{\textbf{9.80}} & 9.16 & 8.51 & 10.44 & 10.21 & 11.30 & 10.66 & \textcolor{red}{\textbf{10.05}} \\
    \midrule
    \multirow{3}{*}{SINDy-AR} & 0.001 & 10.0 & 6.10 & 6.33 & 5.91 & 6.60 & 5.76 & 5.13 & 4.91 & 4.44 & \textcolor{blue}{\textbf{5.65}} & 5.25 & 5.35 & 6.29 & 6.28 & 7.07 & 7.45 & \textcolor{blue}{\textbf{6.28}} \\
     & 0.005 & 0.1 & 5.84 & 6.34 & 6.37 & 7.09 & 6.19 & 5.41 & 5.05 & 4.30 & 5.82 & 5.67 & 5.70 & 6.37 & 5.79 & 6.67 & 7.42 & 6.27 \\
     & 0.01 & 10.0 & 6.35 & 6.65 & 6.24 & 6.89 & 6.04 & 5.58 & 5.35 & 4.50 & 5.95 & 5.61 & 5.71 & 6.55 & 6.25 & 7.12 & 7.50 & 6.46 \\
    \bottomrule
  \end{tabular}
  }
\end{sidewaystable}

\begin{sidewaystable}[htbp]
  \centering
  \caption{COAE-TIEGCM emulation, polynomial of order 2 basis functions. All weather conditions. Mean MAPE (\%) by year for the 5-day propagation window.}
  \label{tab:mape_year_5d_cae_poly2}
  \resizebox{\textwidth}{!}{%
  \begin{tabular}{lllrrrrrrrrr|rrrrrrr}
    \toprule
    \multicolumn{3}{c}{} & \multicolumn{9}{c}{Train} & \multicolumn{7}{c}{Test} \\
    \cmidrule(lr){4-12} \cmidrule(lr){13-19}
    Model & $\lambda$ & $\alpha_{\text{ridge}}$ & 2002 & 2003 & 2004 & 2005 & 2006 & 2007 & 2008 & 2009 & Avg & 1996 & 1997 & 1998 & 1999 & 2000 & 2001 & Avg \\
    \midrule
    F10.7 & --- & --- & 195.0 & 140.0 & 114.5 & 97.7 & 79.8 & 69.1 & 62.9 & 61.2 & 102.5 & 71.7 & 82.1 & 122.4 & 155.1 & 186.8 & 182.2 & 133.4 \\
    Ap & --- & --- & 12.76 & 21.24 & 13.03 & 13.08 & 8.27 & 7.32 & 6.79 & 3.87 & 10.80 & 8.55 & 8.15 & 11.64 & 12.16 & 14.69 & 12.52 & 11.28 \\
    \midrule
    DMDc & --- & --- & 11.40 & 13.48 & 11.88 & 12.18 & 11.21 & 11.60 & 11.94 & 10.17 & \textcolor{red}{\textbf{11.73}} & 11.11 & 10.14 & 12.33 & 12.38 & 12.82 & 12.70 & \textcolor{red}{\textbf{11.91}} \\
    \midrule
    \multirow{3}{*}{SINDy-AR} & 0.001 & 10.0 & 7.67 & 8.23 & 7.90 & 8.44 & 7.58 & 7.44 & 7.26 & 6.00 & \textcolor{blue}{\textbf{7.57}} & 7.02 & 6.82 & 7.45 & 7.65 & 8.79 & 8.36 & \textcolor{blue}{\textbf{7.68}} \\
     & 0.005 & 0.1 & 8.07 & 8.91 & 8.78 & 9.37 & 7.87 & 7.67 & 7.41 & 6.05 & 8.02 & 7.57 & 7.25 & 8.13 & 7.72 & 8.82 & 8.36 & 7.97 \\
     & 0.01 & 10.0 & 7.56 & 8.48 & 8.03 & 8.43 & 7.57 & 7.39 & 7.25 & 5.97 & 7.59 & 6.99 & 6.89 & 7.70 & 7.81 & 9.09 & 8.87 & 7.89 \\
    \bottomrule
  \end{tabular}
  }
\end{sidewaystable}

\begin{sidewaystable}[htbp]
  \centering
  \caption{COAE-TIEGCM emulation, polynomial of order 3 basis functions. All weather conditions. Mean MAPE (\%) by year for the 5-day propagation window.}
  \label{tab:mape_year_5d_cae_poly3}
  \resizebox{\textwidth}{!}{%
  \begin{tabular}{lllrrrrrrrrr|rrrrrrr}
    \toprule
    \multicolumn{3}{c}{} & \multicolumn{9}{c}{Train} & \multicolumn{7}{c}{Test} \\
    \cmidrule(lr){4-12} \cmidrule(lr){13-19}
    Model & $\lambda$ & $\alpha_{\text{ridge}}$ & 2002 & 2003 & 2004 & 2005 & 2006 & 2007 & 2008 & 2009 & Avg & 1996 & 1997 & 1998 & 1999 & 2000 & 2001 & Avg \\
    \midrule
    F10.7 & --- & --- & 195.0 & 140.0 & 114.5 & 97.7 & 79.8 & 69.1 & 62.9 & 61.2 & 102.5 & 71.7 & 82.1 & 122.4 & 155.1 & 186.8 & 182.2 & 133.4 \\
    Ap & --- & --- & 12.76 & 21.24 & 13.03 & 13.08 & 8.27 & 7.32 & 6.79 & 3.87 & 10.80 & 8.55 & 8.15 & 11.64 & 12.16 & 14.69 & 12.52 & 11.28 \\
    \midrule
    DMDc & --- & --- & 11.40 & 13.48 & 11.88 & 12.18 & 11.21 & 11.60 & 11.94 & 10.17 & \textcolor{red}{\textbf{11.73}} & 11.11 & 10.14 & 12.33 & 12.38 & 12.82 & 12.70 & \textcolor{red}{\textbf{11.91}} \\
    \midrule
    \multirow{3}{*}{SINDy-AR} & 0.001 & 10.0 & 6.59 & 6.74 & 6.42 & 7.48 & 6.24 & 5.67 & 5.42 & 4.84 & \textcolor{blue}{\textbf{6.18}} & 5.77 & 5.85 & 6.68 & 6.80 & 8.03 & 8.37 & \textcolor{blue}{\textbf{6.92}} \\
     & 0.005 & 0.1 & 6.22 & 6.68 & 6.81 & 8.24 & 6.88 & 5.94 & 5.55 & 4.68 & 6.38 & 6.27 & 6.34 & 6.78 & 6.21 & 7.31 & 8.23 & 6.86 \\
     & 0.01 & 10.0 & 6.95 & 7.12 & 6.81 & 7.85 & 6.80 & 6.25 & 6.01 & 5.02 & 6.60 & 6.26 & 6.40 & 7.00 & 6.83 & 8.11 & 8.45 & 7.18 \\
    \bottomrule
  \end{tabular}
  }
\end{sidewaystable}

\begin{sidewaystable}[htbp]
  \centering
  \caption{COAE-TIEGCM emulation, polynomial of order 2 basis functions. Mean MAPE (\%) by year for the storm-period propagation window.}
  \label{tab:mape_year_storm_cae_poly2}
  \resizebox{\textwidth}{!}{%
  \begin{tabular}{lllrrrrrrrrr|rrrrrrr}
    \toprule
    \multicolumn{3}{c}{} & \multicolumn{9}{c}{Train} & \multicolumn{7}{c}{Test} \\
    \cmidrule(lr){4-12} \cmidrule(lr){13-19}
    Model & $\lambda$ & $\alpha_{\text{ridge}}$ & 2002 & 2003 & 2004 & 2005 & 2006 & 2007 & 2008 & 2009 & Avg & 1996 & 1997 & 1998 & 1999 & 2000 & 2001 & Avg \\
    \midrule
    F10.7 & --- & --- & 190.0 & 133.4 & 122.5 & 102.1 & 81.6 & 70.0 & 65.8 & 60.0 & 103.2 & 74.0 & 82.4 & 124.9 & 157.1 & 192.0 & 186.8 & 136.2 \\
    Ap & --- & --- & 28.40 & 34.95 & 39.05 & 32.20 & 23.93 & 20.81 & 20.87 & 11.82 & 26.50 & 22.82 & 21.24 & 30.26 & 26.90 & 31.08 & 32.57 & 27.48 \\
    \midrule
    DMDc & --- & --- & 15.96 & 15.31 & 19.01 & 17.45 & 15.82 & 13.96 & 15.16 & 12.33 & \textcolor{red}{\textbf{15.63}} & 18.08 & 14.40 & 16.85 & 13.24 & 16.55 & 16.57 & \textcolor{red}{\textbf{15.95}} \\
    \midrule
    \multirow{3}{*}{SINDy-AR} & 0.001 & 10.0 & 10.32 & 12.41 & 15.50 & 14.45 & 12.26 & 12.53 & 12.26 & 10.92 & \textcolor{blue}{\textbf{12.58}} & 12.78 & 12.24 & 14.27 & 10.89 & 10.90 & 11.99 & \textcolor{blue}{\textbf{12.18}} \\
     & 0.005 & 0.1 & 11.33 & 14.43 & 17.79 & 16.94 & 13.58 & 13.09 & 12.92 & 10.69 & 13.85 & 13.04 & 13.30 & 16.24 & 11.76 & 11.99 & 12.19 & 13.09 \\
     & 0.01 & 10.0 & 10.20 & 13.43 & 15.16 & 15.08 & 12.18 & 11.81 & 12.07 & 10.48 & 12.55 & 11.75 & 11.98 & 14.42 & 10.77 & 11.15 & 11.32 & 11.90 \\
    \bottomrule
  \end{tabular}
  }
\end{sidewaystable}

\begin{sidewaystable}[htbp]
  \centering
  \caption{COAE-TIEGCM emulation, polynomial of order 3 basis functions. Mean MAPE (\%) by year for the storm-period propagation window.}
  \label{tab:mape_year_storm_cae_poly3}
  \resizebox{\textwidth}{!}{%
  \begin{tabular}{lllrrrrrrrrr|rrrrrrr}
    \toprule
    \multicolumn{3}{c}{} & \multicolumn{9}{c}{Train} & \multicolumn{7}{c}{Test} \\
    \cmidrule(lr){4-12} \cmidrule(lr){13-19}
    Model & $\lambda$ & $\alpha_{\text{ridge}}$ & 2002 & 2003 & 2004 & 2005 & 2006 & 2007 & 2008 & 2009 & Avg & 1996 & 1997 & 1998 & 1999 & 2000 & 2001 & Avg \\
    \midrule
    F10.7 & --- & --- & 190.0 & 133.4 & 122.5 & 102.1 & 81.6 & 70.0 & 65.8 & 60.0 & 103.2 & 74.0 & 82.4 & 124.9 & 157.1 & 192.0 & 186.8 & 136.2 \\
    Ap & --- & --- & 28.40 & 34.95 & 39.05 & 32.20 & 23.93 & 20.81 & 20.87 & 11.82 & 26.50 & 22.82 & 21.24 & 30.26 & 26.90 & 31.08 & 32.57 & 27.48 \\
    \midrule
    DMDc & --- & --- & 15.96 & 15.31 & 19.01 & 17.45 & 15.82 & 13.96 & 15.16 & 12.33 & \textcolor{red}{\textbf{15.63}} & 18.08 & 14.40 & 16.85 & 13.24 & 16.55 & 16.57 & \textcolor{red}{\textbf{15.95}} \\
    \midrule
    \multirow{3}{*}{SINDy-AR} & 0.001 & 10.0 & 8.72 & 9.50 & 12.12 & 13.19 & 10.27 & 9.04 & 8.96 & 7.16 & \textcolor{blue}{\textbf{9.87}} & 10.85 & 10.60 & 12.44 & 9.87 & 11.01 & 12.71 & \textcolor{blue}{\textbf{11.25}} \\
     & 0.005 & 0.1 & 8.15 & 8.76 & 12.22 & 14.47 & 11.36 & 9.66 & 9.18 & 7.43 & 10.15 & 12.13 & 11.62 & 12.27 & 8.71 & 9.99 & 12.42 & 11.19 \\
     & 0.01 & 10.0 & 9.82 & 9.53 & 12.41 & 13.40 & 11.20 & 9.44 & 9.24 & 7.82 & 10.36 & 10.63 & 10.97 & 12.55 & 9.49 & 10.74 & 12.64 & 11.17 \\
    \bottomrule
  \end{tabular}
  }
\end{sidewaystable}

\end{document}